\documentclass[showpacs,12pt,a4paper]{article}  

\usepackage{geometry}
\usepackage{cite}
\usepackage{amsmath,amssymb}
\usepackage{latexsym,amscd,amsthm,amsxtra,graphicx,mathrsfs,amsfonts}
\usepackage{verbatim}

\usepackage{dcolumn}
\usepackage[dvipdfm]{hyperref}

\usepackage{color}

\usepackage{multirow}
\usepackage{extarrows}
\usepackage{empheq,mdframed,lipsum}
\usepackage{diagbox}

\usepackage{caption,epsfig,subfig}

\usepackage{dsfont}
\usepackage{array}
\usepackage{booktabs}

\usepackage{xpatch}
\usepackage{blindtext}

\usepackage{bm}
\usepackage{cases}

\usepackage{simplewick}
\usepackage{pdfpages}

\newcommand{\ovl}{\overline} 
\newcommand{\chiP}{\chi_{P}}

\newcommand{\be}{\begin{equation}}
\newcommand{\ee}{\end{equation}}
\newcommand{\ba}{\begin{eqnarray}}
\newcommand{\ea}{\end{eqnarray}}
\newcommand{\bea}{\begin{equation}\left\{ \begin{aligned}}
\newcommand{\eea}{ \end{aligned} \right. \end{equation}}
\newcommand{\bc}{\begin{numcases}  }
\newcommand{\ec}{\end{numcases}   }
\newcommand{\non}{\nonumber}
\newcommand{\nn}{\nonumber}

\newcommand{\ben}{\begin{equation*}}
\newcommand{\een}{\end{equation*}}
\newcommand{\ban}{\begin{eqnarray*} }
\newcommand{\ean}{\end{eqnarray*}}

\allowdisplaybreaks[2]

\UseRawInputEncoding
\begin{document}

\title{Scalar Bound States of $D^{\ast}\bar{D}^{\ast}$ and $B^{\ast}\bar{B}^{\ast}$ in the
Bethe-Salpeter Formalism}

\author{
Rui-Cheng LI\thanks{rui-chengli@163.com}\\
{\small\it School of Physical Sciences, University of Chinese Academy of Sciences, Beijing 100049, China}
}

\date{}

\maketitle

\begin{abstract}
We study the scalar bound states of $D^{\ast}\bar{D}^{\ast}$ and $B^{\ast}\bar{B}^{\ast}$ in the
Bethe-Salpeter
formalism,
with the effective interaction kernel
extracted
from the chiral perturbative theory
and the heavy quark effective theory
in the ladder approximation and the covariant instantaneous approximation.
The results show that, in the scalar case ($J=0$), there can only exist $I = 0$ bound states for parameters in proper range,
while there cannot exist the $I = 1$ bound states in the whole reasonable parameter range,
due to more constraints arising from our definition of the Bethe-Salpeter wavefunction.
\end{abstract}


\thispagestyle{empty}

\newpage
\setcounter{page}{1}

\section{Introduction}

The composite states constituted with a pair of vector fields $D^\ast \bar{D}^\ast$ or $B^\ast \bar{B}^\ast$ have been studied in many methods, such as
in the Bethe-Salpeter (BS) equation formalism for bound states of two vector particles\cite{BSE-DstarDstar-KeHongWei},\cite{BSE-DstarDstar-ChenXiaozhao-2013},\cite{BSE-DstarDstar-ChenXiaozhao-2015},
in the BS equation (BSE) formalism for bound states of four quarks\cite{BSE-four-quark},
in the T-matrix formalism\cite{T-matrix-scheme},
in the Schrodinger equation (SchE)  formalism for bound states of two vector particles\cite{SchE-scheme}\cite{single-term-bad-Lagrangian},
in the SchE formalism for bound states of four chiral constituent quarks\cite{ChQM-scheme},
in the QCD sum-rule formalism\cite{sum-rule-scheme}, or
in the approach of fitting experimental data\cite{fit-data-scheme}, etc.
In the previous studies,
the definitions of the bound states and the effective Lagrangian
taken in the BSE formalism in Ref. \cite{BSE-DstarDstar-KeHongWei},\cite{BSE-DstarDstar-ChenXiaozhao-2013},\cite{BSE-DstarDstar-ChenXiaozhao-2015}
are different from the ones taken in the SchE formalism in Ref. \cite{SchE-scheme},\cite{SchE-scheme-CPC-A-Note},
which will not be convenient
to explore
the dependence of
the results
on the choices of the formalism.
So, compared with  Ref. \cite{BSE-DstarDstar-KeHongWei},\cite{BSE-DstarDstar-ChenXiaozhao-2013},\cite{BSE-DstarDstar-ChenXiaozhao-2015}, what will be new in our work is that,
we will take a new investigation to the bound states of $D^\ast \bar{D}^\ast$ and $B^\ast \bar{B}^\ast$ system in the BSE formalism
by applying the same definitions of the bound states and the effective Lagrangian as the ones in the SchE formalism taken
in Ref. \cite{SchE-scheme}. Moreover, due to the new definitions of the bound states with definite isospin quantum numbers
and
a new
analysis on the Lorentz structure, we will take a new form for the the BS wavefunctions (BSWFs).
Besides, we will take a study on the decay widths of the bound states in the BSE formalism, which were not involved in Ref. \cite{BSE-DstarDstar-KeHongWei},\cite{BSE-DstarDstar-ChenXiaozhao-2013},\cite{BSE-DstarDstar-ChenXiaozhao-2015}.

The remainder of this paper is organized as follows.
In Section II, the BSWF and BSE formalism of two vector particle system will be constructed
as in Ref. \cite{BSE-origin},\cite{DSE-Schwinger},\cite{WuXingHua-KK}.
In Section III, the effective interaction kernel in the BSE,
will be extracted from the chiral perturbative theory (ChPT)
and the heavy quark effective theory (HQET)
in the ladder approximation and the covariant instantaneous approximation.
In Section IV,
we will give the construction of the Lorentz structure of the BSWF.
In Section V, we will show the method of solving the BSE and the normalization of the BSWF.
In Section VI, we will illustrate how to calculate the decay width of the bound states in the BS formalism.
Section VII, some numerical results will be listed.
Finally, the conclusion and an outlook are given in Section VIII.

\section{Bethe-Salpeter Wavefunction and Bethe-Salpeter Equation}

\subsection{The isospin multiplets constituted by $D^{\ast}\bar{D}^{\ast}$ and the Bethe-Salpeter wavefunction}

For the single-charmed mesons $\{D^{\ast}\}$ and their antiparticles $\{\widetilde{D}^{\ast}\}$,
if we take the notations for the particles
as
\be
D^{\ast}_1 =D^{\ast +}, \widetilde{D}^{\ast}_1 =D^{\ast -},\,
D^{\ast}_2 =D^{\ast 0}, \widetilde{D}^{\ast}_2 =\bar{D}^{\ast 0},
\ee
and
define the corresponding field operators as
\ba
D^{\ast\mu}_i(x)
&=& \int {d^3 p\over (2\pi)^3}{1\over \sqrt{2 E_{\bf p}}}
\sum^{3}_{r=1}
\left[  a_{i}       ({\bf p},r)  \varepsilon_{i}^{\mu    }({\bf p},r)   e^{-ip \cdot x}
+       a_{i}^{\dag}({\bf p},r)  \varepsilon_{i}^{\mu \,*}({\bf p},r)   e^{ ip \cdot x}  \right] ,\non\\
\widetilde{D}^{\ast\mu}_i(x)
&=& \int {d^3 p\over (2\pi)^3}{1\over \sqrt{2 E_{\bf p}}}
\sum^{3}_{r=1}
\left[  b_{i}       ({\bf p},r)  \xi_{i}^{\mu    }({\bf p},r)   e^{-ip \cdot x}
+       b_{i}^{\dag}({\bf p},r)  \xi_{i}^{\mu \,*}({\bf p},r)   e^{ ip \cdot x}  \right] ,\, i=1,2 ,  \label{define-Dstar-field}\non\\
\ea
then,
there are two isospin doublets
(we choose the convention
on the isospin multiplets the same as that in Ref. \cite{Perkins}):
\ba
\widetilde{D}^{\ast} &\equiv&
( \widetilde{D}^{\ast}_1 ,   \widetilde{D}^{\ast}_2 )^T, \label{isospin-doublet-1}\\
D^{\ast} &\equiv&
( D^{\ast}_1  ,   - D^{\ast}_2  )^T.
\label{isospin-doublet-2}
\ea
with the superscript $T$ denoting transposition.
Note,  once the doublet in (\ref{isospin-doublet-1}) and the convention of C-parity operation have been defined, for the anti-particle the minus sign in (\ref{isospin-doublet-2}) would automatically arise, which is independent on the details of $\{D^{\ast}\}$ and $\{\widetilde{D}^{\ast}\}$ mesons  (e.g., the flavor wavefunctions) at quark level.
Consequently, the isospin quantum numbers of $D^{\ast} \widetilde{D}^{\ast}$ composite system can be $0$ or $1$;
and, the iso-scalar bound state
can be
written as
\ba
|P\rangle_{0}
= {1\over\sqrt{2}}\left| D^{\ast +}D^{\ast -}  +  D^{\ast 0}\bar{D}^{\ast 0} \right\rangle \,,
\ea
while the three components of the iso-vector states can be
written as
\ba
|P\rangle_{1,0}  &=& {1\over\sqrt{2}} \left| D^{\ast +}D^{\ast -}  - D^{\ast 0}\bar{D}^{\ast 0}  \right\rangle,\non\\
|P\rangle_{1,+1} &=&                  \left| D^{\ast +} \bar{D}^{\ast 0}  \right\rangle,\non\\
|P\rangle_{1,-1} &=&                - \left| D^{\ast 0}   D^{\ast -}   \right\rangle .
\ea
In the following of this paper,
it could be out of doubt that, the notation $D^{\ast}_i$ (or $\widetilde{D}^{\ast}_i$) in a bra (or ket) should be understood as a kind of flavor quantum number
while the one out of a bra (or ket) should be understood as a kind of field operator, respectively. $D^{\ast\mu}_i(x)$ and $\widetilde{D}^{\ast\mu}_i(x)$ are defined as real-valued fields, so the hermite conjugate fields will be $[D^{\ast\mu}_i(x)]^\dag=D^{\ast\mu}_i(x)$ and $[\widetilde{D}^{\ast\mu}_i(x)]^\dag=\widetilde{D}^{\ast\mu}_i(x)$.

Now, we can define a class of hadronic matrix elements, i.e., the so-called Bethe-Salpeter (BS) wave functions,
as
\ba
\chi^{I=0,I_3=0}_{P}(x_1,x_2) &\equiv&
\langle \Omega| {\rm T}\,\{
{1\over\sqrt{2}} [ D^{\ast}_{1}(x_1)\widetilde{D}^{\ast}_{1} (x_2) + D^{\ast}_{2}(x_1)\widetilde{D}^{\ast}_{2}(x_2) ]
\}
|P\rangle_{0,0}  ,\non\\
\chi^{I=1,I_3=0}_{P}(x_1,x_2) &\equiv&
\langle \Omega| {\rm T}\,\{
{1\over\sqrt{2}} [D^{\ast}_{1}(x_1)\widetilde{D}^{\ast}_{1} (x_2) - D^{\ast}_{2}(x_1)\widetilde{D}^{\ast}_{2}(x_2) ]
\}
|P\rangle_{1,0} ,\non\\
\chi^{I=1,I_3=+1}_{P}(x_1,x_2) &\equiv&
\langle \Omega| {\rm T}\,\{
D^{\ast}_{1}(x_1)\widetilde{D}^{\ast}_{2} (x_2)
\}
|P\rangle_{1,+1},\non\\
\chi^{I=1,I_3=-1}_{P}(x_1,x_2) &\equiv&
\langle \Omega| {\rm T}\,\{
D^{\ast}_{2}(x_1)\widetilde{D}^{\ast}_{1} (x_2)
\}
|P\rangle_{1,-1} ,  \label{define-BSWF}
\ea
where $\langle \Omega|$ is the vacuum in the interaction picture. By recalling the fields $D^{\ast\mu}_i(x)$ and $\widetilde{D}^{\ast\mu}_i(x)$ are hermite self-conjugate, i.e., $[D^{\ast\mu}_i(x)]^\dag=D^{\ast\mu}_i(x)$ and $[\widetilde{D}^{\ast\mu}_i(x)]^\dag=\widetilde{D}^{\ast\mu}_i(x)$, we can have
\ba
&& \langle \Omega| {\rm T}\,\{  D^{\ast}_{i}(x_1)\widetilde{D}^{\ast}_{j} (x_2)\} | D^{\ast}_{i} \widetilde{D}^{\ast}_{j} \rangle_{I,I_3}  \non\\
&=&\langle \Omega| {\rm T}\,\{  D^{\ast}_{i}(x_1)\widetilde{D}^{\ast \dag}_{j} (x_2)\} | D^{\ast}_{i} \widetilde{D}^{\ast}_{j} \rangle_{I,I_3}  \non\\
&=&\langle \Omega| {\rm T}\,\{  D^{\ast \dag}_{i}(x_1)\widetilde{D}^{\ast}_{j} (x_2)\} | D^{\ast}_{i} \widetilde{D}^{\ast}_{j} \rangle_{I,I_3}  \non\\
&=&\langle \Omega| {\rm T}\,\{  D^{\ast \dag}_{i}(x_1)\widetilde{D}^{\ast \dag}_{j} (x_2)\} | D^{\ast}_{i} \widetilde{D}^{\ast}_{j} \rangle_{I,I_3}.
\label{hc-BSWF}
\ea
We will not consider the trivial hadronic matrix elements such as
\ba
&&\langle \Omega| {\rm T}\,\{  D^{\ast (\dag)}_{i}(x_1)D^{\ast (\dag)}_{j} (x_2)\} | D^{\ast}_{i} \widetilde{D}^{\ast}_{j}  \rangle_{I,I_3} \non\\
&=&
\langle \Omega| {\rm T}\,\{  \widetilde{D}^{\ast (\dag)}_{i}(x_1)\widetilde{D}^{\ast (\dag)}_{j} (x_2)\} | D^{\ast}_{i} \widetilde{D}^{\ast}_{j} \rangle_{I,I_3}
=0,
\label{trivial-BSWF}
\ea
since we have define the system $|P\rangle_{I,I_3}$ as a $D^{\ast} \widetilde{D}^{\ast}$ composite 2-body system.
Besides, we want to point out that,
definition in the form\cite{BSE-DstarDstar-KeHongWei} of $\chi_P(x_1,x_2) = \langle \Omega| {\rm T}\,\{ \phi_{1\mu}(x_1) \phi^{\mu}_{2}(x_2) \} |P\rangle$ are not corresponding to a generally proper BSWF. \\

By recalling the property of the ladder operator $I^{\pm}$
\be
I^{\pm} |I,I_3\rangle =\sqrt{I(I+1)-I_3(I_3 \pm 1)} |I,I_3\pm 1 \rangle,
\ee
we can know that, $\frac{1}{\sqrt{2}}I^{\pm}$ is a normalized and unitary operator and it can serve as a group element in the isospin $SU(2)$ case.
Thus, we can get
\ba
&&\chi^{I=1,I_3=-1}_{P}(x_1,x_2)  \equiv
\langle \Omega| {\rm T}\,\{
D^{\ast}_{2}(x_1)\widetilde{D}^{\ast}_{1} (x_2)
\}
|P\rangle_{1,-1}\non\\
&=&
\langle \Omega| {\rm T}\,\{
\frac{1}{\sqrt{2}} I^{-} \cdot\frac{1}{\sqrt{2}}I^{+}\cdot(-1) \cdot [  -  D^{\ast}_{2}(x_1)  \widetilde{D}^{\ast}_{1} (x_2)]\frac{1}{\sqrt{2}} I^{-} \cdot \frac{1}{\sqrt{2}}I^{+}
\}
|P\rangle_{1,-1}\non\\
&=&
-\langle \Omega| {\rm T}\,\{
 [ {1\over\sqrt{2}} (D^{\ast +}D^{\ast -}  - D^{\ast 0}\widetilde{D}^{\ast 0}  )]
\}
\cdot |P\rangle_{1,0}\non\\
&=&
-\chi^{I=1,I_3=0}_{P}(x_1,x_2) ,
\ea
and similarly
\ba
\chi^{I=1,I_3=+1}_{P}(x_1,x_2)
=\chi^{I=1,I_3=0}_{P}(x_1,x_2) ,\quad \mbox{etc.}.
\ea
That means,
we can define a common BS wave function
\ba
\chi^{(I)}_{P}(x_1,x_2) \equiv \chi^{I,I_3=0}_{P}(x_1,x_2)    \label{common-BSWF}
\ea
which depends only on the isospin $I$ rather than the $I_3$ component
and
we can write
\be
\chi^{(ij)\mu\kappa }_{(I,I_3)P}(x_1,x_2) \equiv
\langle \Omega| {\rm T}\,\{ D^{\ast\mu}_i(x_1) \widetilde{D}^{\ast\kappa }_j(x_2) \} |P\rangle_{I,I_3}
= C_{(I,I_3)}^{ij}\,\chi^{(I)\mu\kappa }_{P}(x_1,x_2)  , \label{projection-of-BS-state}
\ee
with the isospin coefficients $C^{ij}_{(I,I_3)}$ for the iso-scalar state being
\ba
\label{isoscalar-coefficient-0}
C_{(0,0)}^{11}=C_{(0,0)}^{22} = 1/\sqrt{2}\,,\qquad \hbox{else } = 0 \,,
\ea
and for the iso-vector state being
\ba
\label{isovector-coefficient-1}
C_{(1,0)}^{11}=-C_{(1,0)}^{22} = 1/\sqrt{2}\,,\quad
C_{(1,+1)}^{12} = 1\,,\quad C_{(1,-1)}^{21} = -1
\,,\qquad \hbox{else } = 0 \,.
\ea

We can also define a class of conjugate BS wave-functions as
\ba
\bar{\chi}^{(ij)\mu\kappa }_{(I,I_3)P}(x_1,x_2)
\equiv _{I,I_3}\langle P|T\{ [D_{i}^{\ast\mu}(x_{1}) ]^\dagger[\widetilde{D}_{j}^{\ast\kappa  }(x_{2})]^\dagger\}|\Omega\rangle
=\left[ C_{(I,I_3)}^{ij}\,\chi^{(I)\mu\kappa }_{P}(x_1,x_2) \right]^{\dagger }.\non\\
\label{define-conj-BSWF}
\ea
According to (\ref{BSWF-Fourier}), we can have the orthogonal relations
\be
\int d^4X\chi_{P}\chi_{P'}=0,\,\int d^4X\bar{\chi}_{P}\bar{\chi}_{P'}=0 ,\quad(P\neq P').
\label{BSW-orthogonal-relation}
\ee

\subsection{The Bethe-Salpeter equation and the normalization condition}

\begin{figure}[!htbp]
\centering
\includegraphics[scale=0.6]{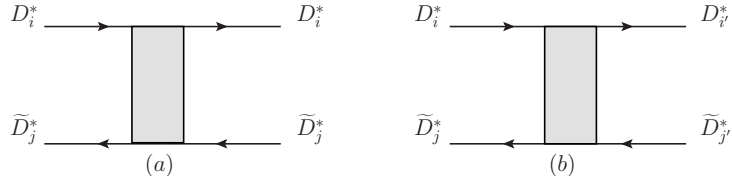}
\caption{Interactions in the system of $D^{\ast}\bar{D}^{\ast}$, with a gray box as an effective vertex.}
\label{Kernel-Dstar-Dstarbar}
\end{figure}

\begin{figure}[!htbp]
\centering
\includegraphics[scale=0.6]{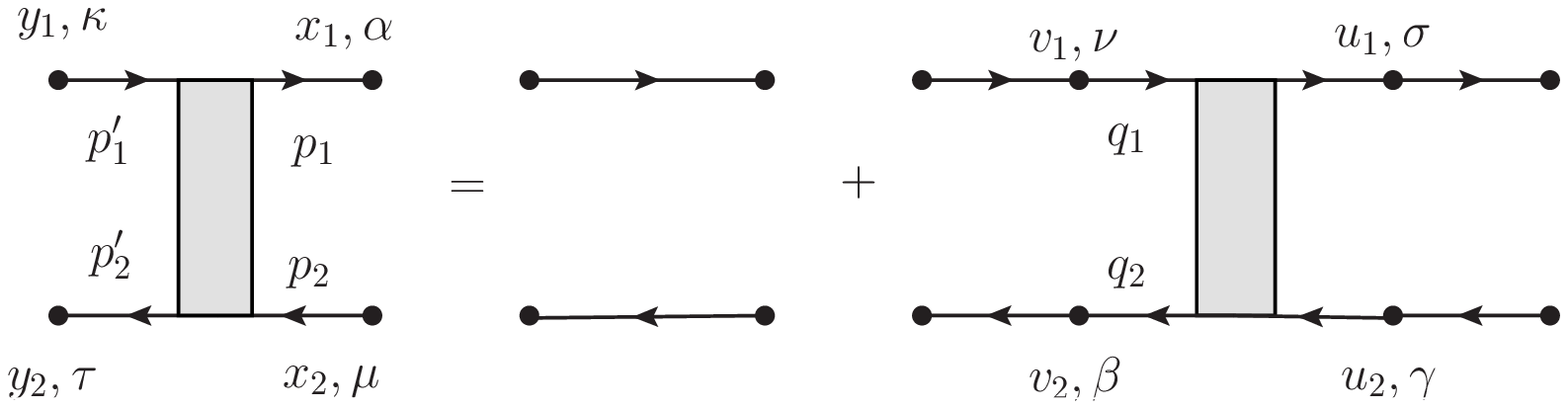}\\
\centering (a)\\
\includegraphics[scale=0.6]{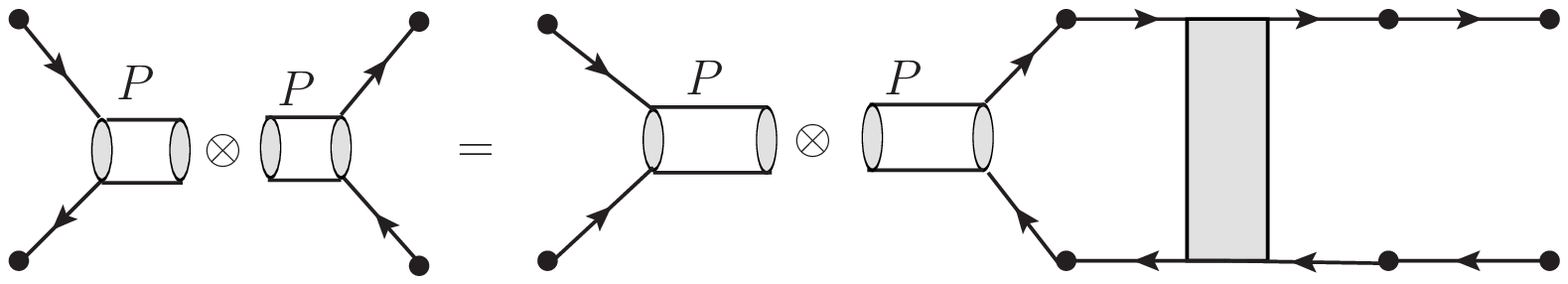}\\
\centering (b)
\caption{Feynman diagrams to represent the Dyson-Schwinger equation (a) and the Bethe-Salpeter equation (b), with a gray box or circle as an effective vertex.}
\label{DSE-and-BSE-Feynman-Diagram}
\end{figure}

Now, only in this subsection, as a simplified version for our derivation, we only consider the interaction shown in Fig.-\ref{Kernel-Dstar-Dstarbar}(a) and ignore the one  in Fig.-\ref{Kernel-Dstar-Dstarbar}(b).
Besides,
we temporarily omit the superscript $(ij)$ and the subscript $(I,I_3)$ for convenience to derive the Bethe-Salpeter equation for the BS wave-function, i.e.,
\ba
\chiP^{\alpha \mu }(x_1,x_2)  \equiv \chi^{(ij)\alpha \mu }_{(I,I_3)P}(x_1,x_2),\,
\bar{\chi}_{P}^{\kappa \tau  }(y_1,y_2) \equiv \bar{\chi}^{(ij)\kappa \tau }_{(I,I_3)P}(y_1,y_2). \label{chi-omit-scripts}
\ea

Define $X$ to be the coordinate of center-of-mass and $x$ to be the relative coordinate of particle $D^{\ast}_i$ and $\widetilde{D}^{\ast}_j$, as
\be
X=\eta_1 x_1 + \eta_2 x_2 , \quad x = x_1 - x_2 ,
\ee
with $\eta_a \equiv m_a/(m_1+m_2)$ and $m_a\, (a=1,2)$ the mass of particle $a$; inversely, there will be
\ba
x_1 = X + \eta_2 x ,  \quad x_2 = X - \eta_1 x  .
\ea
Besides, define $P$ to be the total momentum of the composite system, and $p$ to be a relative momentum, i.e.,
\ba
P = p_1 + p_2  \,,  p = \eta_2p_1 - \eta_1p_2,
\ea
where
$P$ and $p_{1,2}$ are the conjugate momentums of $X$ and $x_{1,2}$, respectively;
inversely, there will be
\ba
p_1 = \eta_1 P + p\,,\quad p_2 = \eta_2 P - p \,.\label{momentum-transform-inverse}
\ea
Note that the momentums $p_{1,2}$ for particles in a bound state $|P\rangle$ are off-shell, i.e., $p_{i}^{2}\neq m_i^2$.

Firstly, as shown in Fig.-\ref{DSE-and-BSE-Feynman-Diagram}, we have
the four-point Green function (in the case of $i=1$ and $j=2$)
\ba
G^{\alpha\mu\kappa \tau }(x_1,x_2;y_1,y_2)
\equiv  \langle \Omega| {\rm T}\,\{D_{1}^{\ast \alpha }(x_{1})\widetilde{D}_{2}^{\ast \mu}(x_{2})
(D_{1}^{\ast \kappa}(y_{1})\widetilde{D}_{2}^{\ast \tau }(y_{2}))^\dag \}|\Omega\rangle \,.
\label{four-point-green-function}
\ea
and the corresponding Dyson¨CSchwinger (DS) equation (DSE)
\be
G=G_0+G_0KG
\ee
or in detail
\ba
&&G^{\alpha\mu\kappa \tau }(x_1,x_2;y_1,y_2)\non\\
&=& G^{\alpha\mu\kappa \tau }_{(0)}(x_1,x_2;y_1,y_2) \nonumber\\
&& +\int d^4u_1 d^4 u_2 d^4 v_1 d^4 v_2   G^{\alpha\mu\sigma\gamma}_{(0)}(x_1,x_2;u_1,u_2) \nn\\
&& \cdot  {\ovl K}_{\sigma\gamma\nu \beta } (u_1,u_2;v_1,v_2)  \cdot G^{\nu \beta  \kappa \tau }(v_1,v_2;y_1,y_2) \,,
\label{DSExy}
\ea
where $\bar{K}$ is the so-called interaction kernel from the irreducible Feynman diagrams to ensure the appearance of $G^{\nu \beta  \kappa \tau }(v_1,v_2;y_1,y_2)$ in the integration to maintain the iterative chain form of DS equation,
and $G_{(0)}$ is the so-called four-point free propagator from the disconnected diagram,
\ba
G^{\alpha\mu\kappa \tau }_{(0)}(x_1,x_2;y_1,y_2)  = \Delta_{1}^{\alpha\kappa }(x_1,y_1)\cdot \Delta_{2}^{\mu\tau }(x_2,y_2) \,,
\label{G0}
\ea
with $\Delta_i(x_i, y_i)$ the full propagator of the particle $i$ $(i=1,2)$,
\ba
\Delta_{i}^{\alpha\kappa }(x, y) = \langle \Omega| {\rm T}\,D_{i}^{\ast \alpha}(x)D_{i}^{\ast \kappa}(y)^\dag|\Omega\rangle
=\int {d^4 k_i \over(2\pi)^4} \, e^{-ik_i\cdot (x-y)}\,S_{i}^{\alpha\kappa}(k_i,m_i) \,.\label{propagator-xy}
\ea
Particularly, in the bound state cases, we get a homogeneous equation
\be
G=G_0KG.  \label{homo-DSE}
\ee

Secondly, according to the inverse Fourier transformations
\ba
\chiP^{\alpha\mu}(x_1,x_2) &=& \frac{1}{(2 \pi)^8} \int d^4 p_1 d^4 p_2\, e^{-(ip_1\cdot x_1 + ip_2\cdot x_2)} \chiP^{\alpha\mu}(p_1,p_2) \non\\
&=& \frac{1}{(2 \pi)^8} \int d^4 \bar{P} d^4 p\, e^{-(i \bar{P} \cdot X + ip \cdot x )} \cdot[ (2 \pi)^4\delta^{(4)}(\bar{P}-P)] \chiP^{\alpha\mu}(p) \non\\
&=& e^{-(i P \cdot X )} \chiP^{\alpha\mu}(x)    , \label{BSWF-Fourier}
\ea
with $\bar{P}= p_1+p_2$ and
\be
\chiP^{\alpha\mu}(x)  \equiv \frac{1}{(2 \pi)^4} \int d^4p \, e^{-ip\cdot x}\chiP^{\alpha\mu}(p)  \,,
\ee
we can always define
\ba
\chi^{(I)\alpha\mu}_{P}(x_1,x_2)\equiv  e^{-i P\cdot X}\chi^{(I)\alpha\mu}_{P}(x),\,
\bar{\chi}^{(I)\alpha\mu}_{P}(x_1,x_2)\equiv  e^{ i P\cdot X}\bar{\chi}^{(I)\alpha\mu}_{P}(x)   .
\ea
Then, in the bound state cases, by inserting a complete set of bound states $\{|P \rangle\}$,
\ba
&&G^{\alpha\mu\kappa \tau }(x_1,x_2;y_1,y_2)\nonumber\\
&=&  \langle \Omega| {\rm T}\,\{D_{1}^{\alpha }(x_{1})\widetilde{D}_{2}^{\mu}(x_{2})(D_{1}^{\kappa  }(y_{1})\widetilde{D}_{2}^{\tau }(y_{2}))^\dag \}|\Omega\rangle \nonumber\\
&=&\sum_{P}   \left[ \langle \Omega| {\rm T}\,\{D_{1}^{\alpha }(x_{1})\widetilde{D}_{2}^{\mu}(x_{2})\} |P\rangle   \right.\nn\\
&& \left. \cdot \frac{1}{2E_{\bf P}}  \langle P|  {\rm T}\,\{ (D_{1}^{\kappa  }(y_{1})\widetilde{D}_{2}^{\tau }(y_{2}))^\dag \}|\Omega\rangle \right]_{\min\{x_1^0,x_2^0\}>\max\{y_1^0,y_2^0\}}  \nn\\
&=& \sum_{P}\left[\frac{1}{2E_{\bf P}} \chiP^{\alpha\mu}(x_1,x_2)\bar{\chi}_{P}^{\kappa  \tau }(y_{1},y_{2}) \right]_{\min\{x_1^0,x_2^0\}>\max\{y_1^0,y_2^0\}} \label{G=chi-dot-chibar}\\
&=& \sum_{P}\left[e^{-iP\cdot (X-Y)}\frac{1}{2E_{\bf P}} \chiP^{\alpha\mu}(x)\bar{\chi}_{P}^{\kappa  \tau }(y) \right]_{\min\{x_1^0,x_2^0\}>\max\{y_1^0,y_2^0\}}   \,.\label{insert-state-non-homo}
\ea
and applying the orthogonal relations (\ref{BSW-orthogonal-relation}), from the homogeneous Dyson-Schwinger equation (\ref{homo-DSE}) we can get the Bethe-Salpeter (BS) equation (BSE) in the coordinate space as
\ba
\chiP^{\alpha\mu}(x_1,x_2)
&=& \int d^4 u_1 d^4 u_2 d^4 v_1 d^4 v_2\, G^{\alpha\mu\sigma\gamma}_{(0)}(x_1,x_2;u_1,u_2)  \nonumber\\
&&\cdot {\ovl K}_{\sigma\gamma\nu \beta } (u_1,u_2;v_1,v_2) \, \chiP^{\nu \beta }(v_1,v_2) \,.
\ea

Thirdly, with the Fourier transformations
\ba
G^{\alpha\mu\kappa \tau }(x_1,x_2;y_1,y_2) &=& \int {d^4P d^4P' d^4p d^4p'\over (2\pi)^{16}}
e^{-iP\cdot X+iP'\cdot Y-ip\cdot x+ip'\cdot y} \, \widetilde G^{\alpha\mu\kappa \tau }(p,p',P,P') \,,\non\\
\label{Green-Fourier}\\
\ovl K_{\sigma\gamma\nu \beta } (x_1,x_2;y_1,y_2) &=& \int {d^4P d^4P' d^4p d^4p'\over (2\pi)^{16}}
e^{-iP\cdot X+iP'\cdot Y-ip\cdot x+ip'\cdot y} \, \ovl K _{\sigma\gamma\nu \beta } (p,p',P,P') \,,\non\\
\ea
and definitions
\ba
\widetilde G^{\alpha\mu\kappa \tau }(p,p',P,P')        &\equiv& (2\pi)^4\delta^4(P-P')\widetilde G^{\alpha\mu\kappa \tau }_P(P,p,p') ,\\
\widetilde G_{(0)}^{\alpha\mu\kappa \tau }(p,p',P,P')  &\equiv& (2\pi)^4\delta^{(4)}(P-P')\widetilde G_{(0)P}^{\alpha\mu\kappa \tau }(P,p,p') \,,\\
{\ovl K}_{\sigma\gamma\nu \beta } (p,p',P,P')      &\equiv& (2\pi)^4\delta^4(P-P'){\ovl K}_{\sigma\gamma\nu \beta }(P,p,p')\,.
\ea
we can get the Dyson-Schwinger equation and the Bethe-Salpeter equation in the momentum space as
\ba
\widetilde G^{\alpha\mu\kappa \tau }_{P}(p,p')&=&
(2\pi)^4 \delta^{4}(p-p')S_{1}^{\alpha\kappa } (p_1,m_1)S_{2}^{\mu\tau }(p_2,m_2) \nonumber\\
&&+
S_{1}^{\alpha\sigma} (p_1,m_1)\int {d^4q  \over (2\pi)^{4}} {\ovl K}_{\sigma\gamma\nu \beta }(P,p,k) \widetilde G^{\nu \beta \kappa \tau }_{P}(P,q,p') S_{2}^{\mu\gamma}(p_2,m_2)\,,
\label{DSE-pq}\\
\chi^{\alpha\mu}_P(p) &=& S_{1}^{\alpha\sigma} (p_1,m_1)\int {d^4q  \over (2\pi)^{4}} {\ovl K}_{\sigma\gamma\nu \beta }(P,p,q) \chi^{\nu \beta }_{P}(k) S_{2}^{\mu\gamma}(p_2,m_2)  \,, \label{BSEpq}
\ea
with
\ba
\widetilde G_{P(0)}^{\alpha\mu\sigma\gamma}(p,q)     =    (2\pi)^4 \delta^{4}(p-q)S_{1}^{\alpha\sigma} (p_1,m_1)S_{2}^{\mu\gamma}(p_2,m_2)  \,.
\ea

Fourthly, as a homogenous integral equation, we need an extra normalization condition to completely solve the Bethe-Salpeter equation with matching the physical boundary conditions.
Indeed, we have the Fourier expansion form
\ba
&& G^{\alpha\mu\kappa \tau }(x_1,x_2;y_1,y_2)  \nonumber\\
&=& \int {d^3{\bf P}\over (2\pi)^3} \,
 e^{-iE_{\bf P}(X^0-Y^0)+i{\bf P}\cdot({\bf X}-{\bf Y})} \frac{1}{2E_{\bf P}}   \chiP^{\alpha\mu}(x) \bar{\chi}_{P}^{\kappa \tau }(y)
\Big|_{\min\{x_1^0,x_2^0\}>\max\{y_1^0,y_2^0\}}  \nonumber\\
&=&  \int {d^3{\bf P}\over (2\pi)^3} \,
 e^{-iE_{\bf P}(X^0-Y^0)+i{\bf P}\cdot({\bf X}-{\bf Y})} \frac{1}{2E_{\bf P}}  \chiP^{\alpha\mu}(x) \bar{\chi}_{P}^{\kappa \tau }(y)
 \nonumber\\
 && \times \theta\left(X^0-Y^0+{\eta_2-\eta_1\over 2}(x^0-y^0)-{|x^0|\over 2}-{|y^0|\over 2}\right)  \label{insert-state-with-theta-function}\\
&=& i \int {\frac{d^4 P}{(2\pi)^4}}e^{i{\bf P}\cdot({\bf X}-{\bf Y})-iP^0(X^0-Y^0)} \frac{1}{2E_{\bf P}}   \chiP^{\alpha\mu}(x) \bar{\chi}_{P}^{\kappa \tau }(y)
\nonumber\\
&&\quad\times\, {1\over P^0-E_{\bf P}+i\epsilon} \,
e^{-i(P^0-E_{\bf P})
\left[(\eta_2-\eta_1)(x^0-y^0)-|x^0|-|y^0|\right]/2}   \,,\label{insert-state-without-theta-function}
\ea
where we compensate a $dP^0$ integration into (\ref{insert-state-with-theta-function}) by using
\be
\theta(x)=  \frac{ i}{2\pi} \int dq^0 \frac{e^{-iq^0x}}{q^0+i\epsilon}  \,,
\ee
thus we can get in the momentum space
\be
\widetilde G_P^{\alpha\mu \kappa \tau } (P,p,p') = {i \over 2E_{\bf P}(P^0-E_{\bf P}+i\epsilon)}\,
\chiP^{\alpha\mu}(p) \bar{\chi}_{P}^{\kappa \tau }(p')
+ ( \hbox{terms regular at $P^0=E_{\bf P}$} ) \,.\label{Green-pq-Fourier}
\ee
After inserting (\ref{Green-pq-Fourier}) into (\ref{DSE-pq}),
and multiplying the factor $\int \frac{d^4p}{(2\pi)^4} \bar{\chi}(p)$  in both sides of (\ref{DSE-pq}),
we can have
\ba
&&\int {d^4q d^4p \over (2\pi)^{8}}
\bar{\chi}_{P}^{\sigma\gamma }(p)
\cdot \left[I_{\sigma\gamma\nu \beta }(P,p,q)
   -    {\ovl K}_{\sigma\gamma\nu \beta }(P,p,q)\right] \nn\\
&& \cdot
\Big[ {i \over 2E_{\bf P}(P^0-E_{\bf P}+i\epsilon)}\,
\chiP^{\nu \beta }(q) \bar{\chi}_{P}^{\kappa \tau }(p')
+ ( \hbox{terms regular at $P^0=E_{\bf P}$} )   \Big]   \nn\\
&=&
\bar{\chi}_{P}^{\sigma\gamma }(p')
\delta_{\sigma}^{\kappa } \delta_{\gamma}^{\tau }   \,,
\ea
with the definition
\be
I_{\sigma\gamma\nu \beta }(P,p,q) \equiv (2\pi)^4 \delta^{4}(q-p) \cdot S_{1 \nu \sigma}^{-1} (p_1,m_1)   S_{2 \beta \gamma}^{-1} (p_2,m_2)    \,.\label{define-I-tensor}
\ee
Particularly, in the limit of $P^0  \rightarrow E_{\bf P}$,
there will only be contribution
from the singular term of (\ref{Green-pq-Fourier}), so we can get
\ba
&& i \int {d^4q d^4p \over (2\pi)^{8}}
\bar{\chi}_{P}^{\sigma\gamma }(p)
\cdot
\frac{\partial}{\partial P^0}\left[ I_{\sigma\gamma\nu \beta }(P,p,q)
   -    {\ovl K}_{\sigma\gamma\nu \beta }(P,p,q)\right]  \cdot
 {1 \over 2E_{\bf P}}\,
 \chiP^{\nu \beta }(q) \bar{\chi}_{P}^{\kappa \tau }(p')   \nn\\
&=&
\bar{\chi}_{P}^{\sigma\gamma }(p')
\delta_{\sigma}^{\kappa } \delta_{\gamma}^{\tau }
=\bar{\chi}_{P}^{\kappa \tau }(p')  \,,\, ( P^0=E_{\bf P})  .
\ea
Thus, by recalling the orthogonal relations (\ref{BSW-orthogonal-relation}), at last we can get the normalization condition of the BS wave-function as
\be
i  \int {d^4q d^4p \over (2\pi)^{8}}
\bar{\chi}_{P}^{\sigma\gamma }(p)
\cdot
\frac{\partial}{\partial P^0}\left[
I_{\sigma\gamma\nu \beta }(P,p,q)
   -    {\ovl K}_{\sigma\gamma\nu \beta }(P,p,q)\right]  \cdot
 {1 \over 2E_{\bf P}}\,
 \chiP^{\nu \beta }(q)
=   1   \,,\,  ( P^0=E_{\bf P}) \,.
\ee

\subsection{The full BSE}

From now on,
as a complete version for our derivation, we consider both the interaction shown in Fig.-\ref{Kernel-Dstar-Dstarbar}(a) and the one  in Fig.-\ref{Kernel-Dstar-Dstarbar}(b).
Besides, we restore the notations
\ba
\chiP^{\alpha\mu}(x_1,x_2)  &\rightarrow& \chi^{(ij)\alpha\mu}_{(I,I_3)P}(x_1,x_2)
= C_{(I,I_3)}^{ij}\,\chi^{(I)\alpha\mu}_{P}(x_1,x_2),\,\non\\
\bar{\chi}_{P}^{\kappa \tau }(y_1,y_2) &\rightarrow& \bar{\chi}^{(ij)\kappa \tau }_{(I,I_3)P}(y_1,y_2)
=\left[ C_{(I,I_3)}^{ij}\,\chi^{(I)\kappa \tau }_{P}(y_1,y_2) \right]^{\dagger },
\ea
as defined in (\ref{projection-of-BS-state},\ref{define-conj-BSWF},\ref{chi-omit-scripts}).
The assignment for the indices and the momentum variables in the BS equation will be set as shown in Fig.-\ref{DSE-and-BSE-Feynman-Diagram} (a).

Due to the isospin symmetry, or the relations in (\ref{common-BSWF}),
we can get the BS equation
for a BS wave-function defined with a state of isospin quantum number $(I,I_3)$  and the operatorsof flavor quantum number $(ij)$
in the coordinate space
\ba
C_{(I,I_3)}^{ij}  \chi_{P}^{(I) \alpha\mu}(x_1,x_2)
&=&\int d^4u_1 d^4 u_2 d^4 v_1 d^4 v_2\, G^{\alpha\mu\sigma\gamma}_{(0)}(x_1,x_2;u_1,u_2) \nonumber\\
&& \cdot \sum_{kl} \left[ {\ovl K}^{ij,kl}_{\sigma\gamma\nu\beta} (u_1,u_2;v_1,v_2) C_{(I,I_3)}^{kl}    \right]
\chi_{P}^{(I) \nu\beta}(v_1,v_2) ,
\label{BSExy}
\ea
that means, a nontrivial $\chi_{P}^{(I) \alpha\mu}(x_1,x_2)$ can only exclusively exist for either the $I=0$ case or the $I=1$ case for a unique set of parameters, due to the different forms of the kernel $\overline{K}$ in the two cases.
Correspondingly, in the momentum space we can get
\be
C_{(I,I_3)}^{ij}  \chi_P^{(I) \alpha\mu}(p)
 =  S_{1}^{\alpha\sigma} (p_1,m_1) S_{2}^{\mu\gamma}(p_2,m_2)
\int {d^4q  \over (2\pi)^{4}}
\sum_{kl} \left[  {\ovl K}^{ij,kl} _{\sigma\gamma\nu\beta}(P,p,q)    C_{(I,I_3)}^{kl}  \right]
\chi_{P}^{(I) \nu\beta}(q)   .   \label{BSEpq}
\ee

Particularly,
according to (\ref{isoscalar-coefficient-0},\ref{isovector-coefficient-1}), after inserting $ij=11$, $I=0, I_3=0$ and $kl=11,22$ into (\ref{BSEpq}) for the $I=0$ case,
or inserting $ij=11$, $I=1, I_3=0$ and $kl=11,22$  into (\ref{BSEpq}) for the $I=1$ case, with
\ba
C_{(0,0)}^{11}=C_{(0,0)}^{22} = 1/\sqrt{2} , \quad
C_{(1,0)}^{11}=-C_{(1,0)}^{22} = 1/\sqrt{2} ,
\ea
there will be
\ba
&&   \chi_P^{(I) \alpha\mu}(p) \nonumber\\
&=&   S_{1}^{\alpha\sigma} (p_1,m_1)  S_{2}^{\mu\gamma}(p_2,m_2)
\int {d^4 q  \over (2\pi)^{4}}
  {\ovl K}^{total} _{\sigma\gamma\nu\beta}(P,p,q)
\chi_{P}^{(I) \nu\beta}(q)   ,   \label{BSE-isospin}
\ea
with
\ba
  {\ovl K}^{total} _{\sigma\gamma\nu\beta}(P,p,q)
\equiv
\left[  {\ovl K}^{11,11} _{\sigma\gamma\nu\beta}(P,p,q)
\pm {\ovl K}^{11,22} _{\sigma\gamma\nu\beta}(P,p,q)      \right],\label{K-total}
\ea
where in (\ref{K-total}) the sign ``$+$" for $I=0$ case and the sign ``$-$"  for $I=1$ case, respectively;
there will also be the normalization condition
\ba
&&i  \int {d^4q d^4p \over (2\pi)^{8}}
\frac{1}{\sqrt{2}  }  \bar{\chi}_{P}^{(I)\sigma\gamma }(p)
\cdot
\frac{\partial}{\partial P^0}\left[
I_{\sigma\gamma\nu\beta}(P,p,q)   -
 {\ovl K}^{total} _{\sigma\gamma\nu\beta}(P,p,q)
\right]  \non\\
&&\cdot
 {1 \over 2E_{\bf P}}\cdot
\frac{1}{\sqrt{2}  } \chi_P^{(I)\nu\beta}(q)
=   1   \,,\,  ( P^0=E_{\bf P})  . \label{BSE-norm-isospin}
\ea

\section{Lagrangian and Effective Kernel}\label{Section-Lagrangian}

\subsection{Lagrangian from ChPT and HQET}

The Lagrangian will be used in our calculation is from combining the chiral perturbative theory (ChPT) and the heavy quark effective theory (HQET), where the single-heavy flavored mesons were treated as matter fields and the light flavored mesons were treated as media fields\cite{HeJun-Lagrangian}.

For the  exchanged light flavored mesons,
the octet pseudoscalar and nonet vector
meson matrices are defined as
\begin{eqnarray}
\mathbb{P}&=&\left(\begin{array}{ccc}
\frac{\pi^{0}}{\sqrt{2}}+\frac{\eta}{\sqrt{6}}&\pi^{+}&K^{+}\\
\pi^{-}&-\frac{\pi^{0}}{\sqrt{2}}+\frac{\eta}{\sqrt{6}}&
K^{0}\\
K^- &\bar{K}^{0}&-\frac{2\eta}{\sqrt{6}}
\end{array}\right),\nonumber\\
\mathbb{V}&=&\left(\begin{array}{ccc}
\frac{\rho^{0}}{\sqrt{2}}+\frac{\omega}{\sqrt{2}}&\rho^{+}&K^{*+}\\
\rho^{-}&-\frac{\rho^{0}}{\sqrt{2}}+\frac{\omega}{\sqrt{2}}&
K^{*0}\\
K^{*-} &\bar{K}^{*0}&\phi
\end{array}\right),\label{vector}
\end{eqnarray}
so, there will be
\ba
\mathbb{P}^\dag  =  \mathbb{P},  (\mathbb{P}_{ab})^\dag  =  \mathbb{P}_{ba},\quad
\mathbb{V}^\dag  =  \mathbb{V},  (\mathbb{V}_{ab})^\dag  =  \mathbb{V}_{ba}.
\ea
Besides, we define $\sigma$ as the only scalar meson mediating interactions between
the single-heavy flavor mesons.

As defined in (\ref{define-Dstar-field}), for the pseudoscalar particles $D$ and $\widetilde{D}$ (or, $B$ and $\widetilde{B}$),
\be
D_1 =D^{ +}, \widetilde{D}_1 =D^{ -},\,
D_2 =D^{ 0}, \widetilde{D}_2 =\bar{D}^{ 0},
\ee
here we also define two real-valued fields as
\ba
D_i(x) &=& \int {d^3 p\over (2\pi)^3}{1\over \sqrt{2 E_{\bf p}}}
 (a_{{\bf p}} \, e^{-ip \cdot x}+ a_{{\bf p}}^{\dag}\,   e^{ip \cdot x}),\non\\
\widetilde{D}_i(x) &=& \int {d^3 p\over (2\pi)^3}{1\over \sqrt{2 E_{\bf p}}}
  (b_{{\bf p}}\,   e^{-ip \cdot x}+ b_{{\bf p}}^{\dag}\,  e^{ip \cdot x}),  \label{define-D-field}
\ea
so,  like $[D^{\ast\mu}_i(x)]^\dag=D^{\ast\mu}_i(x)$ and $[\widetilde{D}^{\ast\mu}_i(x)]^\dag=\widetilde{D}^{\ast\mu}_i(x)$,
the hermite conjugate fields will be $[D_i(x)]^\dag=D_i(x)$ and $[\widetilde{D}_i(x)]^\dag=\widetilde{D}_i(x)$.

From now on, in writing the Lagrangian terms, we will
let $P$ represent the single-heavy flavored meson fields  $P=(D^0,D^+,D_s^+)$ or $P=(B^-, B^0,B_s^0)$,
let $P^*$ represent $P^*=(D^{*0}, D^{*+},D_s^{*+})$ or $P^*=(B^{*-},B^{*0},B_s^{*0})$,
let $\widetilde{P}$ represent the corresponding single-heavy anti-meson field
$\widetilde{P}=(\bar{D}^0,D^-,D_s^-)$ or $\widetilde{P}=(B^+, \bar{B}^0,\bar{B}_s^0)$, and
let  $\widetilde{P}^*$ represent $\widetilde{P}^*=(\bar{D}^{*0},D^{*-},D_s^{*-})$ or $\widetilde{P}^*=(B^{*+}, \bar{B}^{*0},\bar{B}_s^{*0})$;
the interaction Lagrangian terms are listed as below:

\noindent (1) the $P$-$P$-coupled terms\\

\ba
\mathcal{L}_{PP\mathbb{V}}&=&
+i\frac{g_\beta g_v}{\sqrt{2}} P_b \mathbb{V}^{\mu}_{ba}\partial_{\mu}P^{\dag}_{a}
-i\frac{g_\beta g_v}{\sqrt{2}}\partial_{\mu}P_a \mathbb{V}^{\mu}_{ab}P^{\dag}_{b} ,\non\\
\mathcal{L}_{\widetilde{P}\widetilde{P}\mathbb{V}}&=&
+i\frac{g_\beta g_v}{\sqrt{2}} \widetilde{P}^{\dag}_{a} \mathbb{V}^{\mu}_{ab}\partial_{\mu}\widetilde{P}_b
-i\frac{g_\beta g_v}{\sqrt{2}}\partial_{\mu}\widetilde{P}^{\dag}_{b} \mathbb{V}^{\mu}_{ba}\widetilde{P}_a,    \label{L-P-P-V}
\ea

\ba
\mathcal{L}_{PP\sigma}  =
-2g_\sigma m_{P}P_a^\dag P_a\sigma, \quad
\mathcal{L}_{\widetilde{P}\widetilde{P}\sigma}  =
-2g_\sigma m_{P}\tilde{P}_b^\dag \tilde{P}_b\sigma,   \label{L-P-P-Sigma}
\ea

\noindent (2) the $P^{\ast}$-$P^{\ast}$-coupled terms\\

\ba
\mathcal{L}_{P^{\ast}P^{\ast}\mathbb{P}}&=&
+ \frac{g_\pi}{f_\pi} \epsilon_{\mu\nu\alpha\beta} P^{*\nu\dag}_{a} \partial^{\beta}P^{*\mu}_b  \partial^{\alpha}\mathbb{P}_{ba}
+ \frac{g_\pi}{f_\pi}\epsilon_{\mu\nu\alpha\beta} \partial^{\beta} P^{*\mu\dag}_{b}  P^{*\nu}_a  \partial^{\alpha}\mathbb{P}_{ab} ,\non\\
\mathcal{L}_{\widetilde{P}^{\ast}\widetilde{P}^{\ast}\mathbb{P}}&=&
+\frac{g_\pi}{f_\pi} \epsilon_{\mu\nu\alpha\beta} \widetilde{P}^{*\nu}_b \partial^{\beta}\widetilde{P}^{*\mu\dag}_{a} \partial^{\alpha}\mathbb{P}_{ab}
+\frac{g_\pi}{f_\pi} \epsilon_{\mu\nu\alpha\beta} \partial^{\beta}\widetilde{P}^{*\mu}_a \widetilde{P}^{*\nu\dag}_{b} \partial^{\alpha}\mathbb{P}_{ba},
\label{L-Pstar-Pstar-Pion}
\ea
\ba
\mathcal{L}_{P^{\ast}P^{\ast}\mathbb{V}}&=&
- i\frac{g_\beta g_v}{\sqrt{2}} P^{*\nu}_b\mathbb{V}^{\mu}_{ba}\partial_{\mu}P^{*\dag}_{\nu a}
+ i\frac{g_\beta g_v}{\sqrt{2}} \partial_{\mu}P^{*\nu}_a \mathbb{V}^{\mu}_{ab}P^{*\dag}_{\nu b}\nonumber\\
&&-i2\sqrt{2}g_\lambda g_v \bar{M^*}P^{*\mu}_b   (\partial_{\mu}\mathbb{V}_{\nu}-\partial_{\nu}\mathbb{V}_{\mu})_{ba}    P^{*\nu\dag}_a   ,\non\\
\mathcal{L}_{\widetilde{P}^{\ast}\widetilde{P}^{\ast}\mathbb{V}}&=&
-i\frac{g_\beta g_v}{\sqrt{2}}\widetilde{P}^{*\dag}_{\nu a} \mathbb{V}^{\mu}_{ab}\partial_{\mu} \widetilde{P}^{*\nu}_b
+i\frac{g_\beta g_v}{\sqrt{2}}\partial_{\mu}\widetilde{P}^{*\dag}_{\nu b} \mathbb{V}^{\mu}_{ba} \widetilde{P}^{*\nu}_a
\nonumber\\
&&-i2\sqrt{2}g_\lambda g_v \bar{M^*}  \widetilde{P}^{*\mu\dag}_a (\partial_{\mu}\mathbb{V}_{\nu}-\partial_{\nu}\mathbb{V}_{\mu})_{ab}  \widetilde{P}^{*\nu}_b ,\label{L-Pstar-Pstar-V}
\ea
\ba
\mathcal{L}_{P^*P^*\sigma}  =
+2g_\sigma m_{P^*}P_a^{*\dag} P^*_a\sigma ,
\mathcal{L}_{\widetilde{P}^*\widetilde{P}^*\sigma}  =
+2g_\sigma m_{P^*} \tilde{P}_b^{*\dag} \tilde{P}^*_b \sigma ,
\label{L-Pstar-Pstar-Sigma}
\ea

\noindent (3) the $P^{\ast}$-$P$-coupled terms\\

\ba
\mathcal{L}_{P^*P\mathbb{P}} &=&
-\frac{2g_\pi\sqrt{m_Pm_{P^*}}}{f_\pi}
(P_bP^{*\dag}_{a\lambda}\partial^\lambda{}\mathbb{P}_{ba}+ P^*_{a\lambda}P^\dag_{b} \partial^\lambda{}\mathbb{P}_{ab} ),
\nonumber\\
\mathcal{L}_{\widetilde{P}^*\widetilde{P}\mathbb{P}} &=&
+\frac{2g_\pi\sqrt{m_Pm_{P^*}}}{f_\pi}
( \tilde{P}^\dag_{a}\tilde{P}^*_{b\lambda}\partial^\lambda\mathbb{P}_{ab}+\tilde{P}^{*\dag}_{b\lambda}\tilde{P}_a\partial^\lambda\mathbb{P}_{ba} ),
\label{L-Pstar-P-Pion}
\ea

\ba
\mathcal{L}_{P^{*}P  \mathbb{V}}&=&
-i\sqrt{2}g_\lambda g_v\epsilon_{\lambda\alpha\beta\mu}  P^{*\mu\dag}_{a} \partial^{\lambda}P_b   \partial^{\alpha}\mathbb{V}^{\beta}_{ba}
+  i\sqrt{2}g_\lambda g_v \epsilon_{\lambda\alpha\beta\mu}   \partial^{\lambda}P^{* \dag \mu}_a  P_{b} \partial^{\alpha}\mathbb{V}^{\beta}_{ba}
\nonumber\\
&&
+  i\sqrt{2}g_\lambda g_v \epsilon_{\lambda\alpha\beta\mu}   \partial^{\lambda}P^{\dag}_{b}  P^{*\mu}_a \partial^{\alpha} \mathbb{V}^{\beta}_{ab}
-i\sqrt{2}g_\lambda g_v \epsilon_{\lambda\alpha\beta\mu} P^{\dag}_{b}   \partial^{\lambda} P^{*\mu}_a \partial^{\alpha}\mathbb{V}^{\beta}_{ab} ,\nonumber\\
\mathcal{L}_{\widetilde{P}^{*}\widetilde{P}  \mathbb{V}}&=&
+i\sqrt{2}g_\lambda g_v \epsilon_{\lambda\alpha\beta\mu}  \partial^{\lambda}\widetilde{P}^{\dag}_{a} \widetilde{P}^{*\mu}_b \partial^{\alpha}\mathbb{V}^{\beta}_{ab}
- i\sqrt{2}g_\lambda g_v \epsilon_{\lambda\alpha\beta\mu}\widetilde{P}^{\dag}_{a}\partial^{\lambda} \widetilde{P}^{*\mu}_b \partial^{\alpha} \mathbb{V}^{\beta}_{ab}
\nonumber\\
&&
- i\sqrt{2}g_\lambda g_v \epsilon_{\lambda\alpha\beta\mu}\widetilde{P}^{* \dag \mu}_b \partial^{\lambda}\widetilde{P}_{a} \partial^{\alpha}\mathbb{V}^{\beta}_{ba}
+i\sqrt{2}g_\lambda g_v\epsilon_{\lambda\alpha\beta\mu} \partial^{\lambda} \widetilde{P}^{*\mu\dag}_{b}\widetilde{P}_a \partial^{\alpha}V^{\beta}_{ba} ,
\label{L-Pstar-P-V}
\ea

\noindent where
the values of the couplings are chosen as
$g_\pi=0.59$, $f_{\pi}=132 MeV$,
$g_v=5.8$, $g_\beta=0.9$,$g_\lambda=0.56 GeV^{-1}$,
$g_\sigma = g'_\pi/(2\sqrt{6})$,$g'_{\pi}=3.73$ \cite{SchE-scheme},\cite{SchE-scheme-CPC-A-Note},\cite{HeJun-Lagrangian},\cite{HeJun-Lagrangian-couplings}.
Note the $g_{\pi} = 0.59$ here is corresponding to the $g$ in  \cite{SchE-scheme-CPC-A-Note}.
The masses of particles are taken as in Ref. \cite{PDG-2020}.

Note the reversal correspondence relation between the subscript indices of $P^{(*)}$ and $D^{(*)}$, that is, $P^{(*)}_1 = D^{(*)}_2 $ and $P^{(*)}_2 = D^{(*)}_1 $. In all the later sections, the indices $11$ or $22$ in the interaction kernels and amplitudes will be all for the indices $ij$ in $D^{(*)}_i$ and $\widetilde{D}^{(*)}_i$, while the indices of $ab$ in the Lagrangian terms will be not apparently written out.
As noted in (\ref{isospin-doublet-2}),
once the triplet of $P^{(\ast)}$ is defined in the ${\bm 3}$ representation of the $SU(3)_f$ group,
the form of $\widetilde{P}^{(\ast)}$ triplet would automatically be defined in the $\bar{\bm 3}$ representation of the $SU(3)_f$ group,
which is also independent on the details of $D^{\ast}(B^{\ast})$ and $\widetilde{D}^{\ast}(\widetilde{B}^{\ast})$ mesons (e.g., the flavor wavefunctions) at quark level.
By combining that the octets of $\mathbb{P}$ and $\mathbb{V}$ have been defined in the ${\bm 8}$ representation of the $SU(3)_f$ group, the
construction
of the chiral effective Lagrangian is straightforward.

Due to the C-parity symmetry of the Lagrangian, there exists a correspondence
between the terms of mesons $P$ (or $P^*$) and the ones of their antiparticles $\widetilde{P}$ (or $\widetilde{P}^*$)
by a replacement
\ba
&&a\to b  \,,\;  b\to a   \,, \nonumber \\
&&P^{*}_{\mu}\to \widetilde{P}^{*\dag}_\mu,\; P \to -\widetilde{P}^\dag   \,,\nonumber\\
&&P^{*\dag}_\mu\to \widetilde{P}^{*}_\mu,\; P^\dag\to - \widetilde{P}    \,.
\ea

The propagator of $P^{\ast}$ or $\widetilde{P}^{\ast}$ fields in the momentum space, see (\ref{propagator-xy}), is chosen to the form in unitary gauge
as
\ba
S_{i}^{\alpha\sigma}(p_i,m_i)  \equiv
\frac{-i(g^{\alpha\sigma}- p_i^{\alpha}p_i^{\sigma}/m_i^2 )}{p_i^2-m_i^2}
= \mathbb{S}_{i \alpha\sigma}  (p_i,m_i) \cdot  \mathfrak{S}_{i}  (p_i,m_i), \label{propagator}
\ea
with the notations $\mathbb{S}_{\alpha\sigma}$ and $\mathfrak{S}$ defined as the numerator and the other part of the propagator, as
\ba
\mathbb{S}_{i \alpha\sigma}  (p_i,m_i) \equiv -i(g^{\alpha\sigma}-\frac{p_i^{\alpha}p_i^{\sigma}}{m_i^2}),\,
\mathfrak{S}_{i }  (p_i,m_i) \equiv  \frac{1}{p_i^2-m_i^2} . \label{propagator-00}
\ea
That means,
in our procedure of configuring the effective field theory from the HQET,
we only employ
the heavy quark flavor symmetry (HQFS) and the heavy quark spin symmetry (HQSS) onto the couplings in the interaction Lagrangian in
(\ref{L-P-P-V}-\ref{L-Pstar-P-V}), without
transforming the propagator of $P^{\ast}$ or $\widetilde{P}^{\ast}$ into the heavy quark limit.
Besides,
by taking the unitary gauges, the gauge boson propagator contains exactly the three spacelike polarization states, and the unphysical degrees of freedom disappear from the theory. We know the full $S$ matrix is independent of the choice of the gauge fixing conditions, however, the Green functions and the hadronic matrix elements of the BSWF are dependent the gauge choice.

Now we can write out the interaction kernel in the BS equation.
In Fig. \ref{Kernel-Dstar-Dstarbar},
the momentum of exchanged light-flavor meson can be written as
\be
k=p_1-q_1=p-q . \label{define-k}
\ee
In the so-called ladder approximation (i.e., only considering the one-particle-exchange Feynman diagrams), after setting $ij=11$ by following (\ref{BSE-isospin}), we can get
the total interaction kernel by combining contributions from each light flavor meson.
The effective kernel in BSE are listed Appendix \ref{appendix-kernel-in-BSE}.

\subsection{Regular factors}

With the propagators of $D^{\ast} \widetilde{D}^{\ast}$ in (\ref{propagator}) and the interaction kernels in (\ref{Kpion1111}-\ref{Ksigma1122VSKsigma1111}),
in the viewpoint of superficial degree of divergence, now we apparently  write out the power counting form of the BSE (\ref{BSE-isospin}) on variables $p$, $q$ and $k$, as
\ba
\chi_P^{\alpha\mu}(p)
\thicksim \frac{ p^{\alpha}p^{\sigma}}{p^2}  \frac{ p^{\mu} p^{\gamma} }{p^2}
\int d^4 q   \left[\frac{ (p+k) (q+k) k^2 }{k^2 - m_{\phi}^2} \right]_{\sigma\gamma\nu\beta}
\chi_{P}^{\nu\beta}(q)   , \label{BSE-dimension-analysis}
\ea
where $k$  and $m_\phi$ are the momentum and the mass of the exchanged light mesons, respectively.
Although it is not necessary in principle for an effective theory of the low energy hadronic interaction to be renormalizable, it should be at least  renormalizable/convergent to definite perturbative order to practically do calculations.

Here in (\ref{BSE-dimension-analysis}) there are two kinds of
convergence are needed:
one is
the convergence of the BSE depending on the integration $\sim\int d^4 q \overline{K} \cdot \chi_{P}(q) $,
the other is
the convergence of the BSWF depending on the normalization integration $\sim\int d^4 p \bar{\chi}_P(p)\cdot \chi_P(p) $.
Firstly,
after
a factor\cite{FormFactor}
\be
F(k,\Lambda,m_\phi) =  \frac{\Lambda^2 - m_\phi^2}{\Lambda^2 + {\bm k}^2}
\xrightarrow{ (m_\phi  \rightarrow 0 )}  \frac{1}{1  +    {\bm k}^2/\Lambda^2}  ,\quad k=p-q,
\label{define-FF-cutoff}
\ee
or more exactly,  $[ F(k,\Lambda,m_\phi) ]^2$, was introduced into $\overline{K}$ for the $t$-channel processes,
the convergence of the integration $\sim\int d^4 q \overline{K} \cdot \chi_{P}(q) $ can be preserved if
an extra suppressed factors with power of
$\frac{1}{p^m\cdot q^m}$ with $m$ a large enough positive integer
was introduced into the kernel $\overline{K}$ in the BSE (\ref{BSE-isospin}),
where (a) the variables $p$ and $q$ should be given equal status because the constituents into the kernel and out of the kernel are the same ones,
(b) we can get $m>5$ if we treat $k$ is independent on $p$ and $q$, that is not the truth but it is somehow reasonable because $k=p-q$ is independent on $p+q$.
In (\ref{define-FF-cutoff}),
$\Lambda$ is a tunable parameter which could be seemed as a typical energy scale of the interaction, generally taken to be of the order of $\Lambda_{\chi}\simeq4\pi f_\pi$ in the ChPT.
Secondly,
the convergence of the normalization integration $\sim\int d^4 p \bar{\chi}_P(p)\cdot \chi_P(p) $  can be preserved if
there was $\chi_P(p) \sim \frac{1}{p^n} $ with $n>2$, which is coincidentally equivalent to $m>3$
above by inserting $\chi_P(p) \sim \frac{1}{p^n} $ into (\ref{BSE-dimension-analysis}).
In a word, extra suppressed factor is needed to preserve the convergence of both the BSE and the BSWF.

The factor $F(k,\Lambda,m_\phi)$ is also called the (monopole type) form factor (F.F.) of vertex $P^{\ast} \widetilde{P}^{\ast} \phi$ in the Feynman diagrams to characterize the non-point property of the particles $P^{\ast}$ and $\widetilde{P}^{\ast}$ when they are probed by the light mesons $\phi$.
However, to completely renormalize a vertex to ensure the convergence of the physical transition amplitudes, a complete form factor should be defined with all the variables of $p$, $q$ and $k$, or, in other words, to completely characterize the non-point property of a particle $P^{\ast}$ (or, essentially an nonlocal effective vertex), it should be described by all the signals probed by $P^{\ast}$, $\widetilde{P}^{\ast}$ and $\phi$.

Besides of the kernel $\overline{K}$ defined
by the Lagrangian terms in (\ref{L-P-P-V}-\ref{L-Pstar-P-V}), to ensure the convergence of the amplitude of a $P^{\ast} \widetilde{P}^{\ast}\rightarrow P^{\ast} \widetilde{P}^{\ast}$ scattering process,
there is another divergence needed to be suppressed by the factor $\mathcal{F}(k,\Lambda;p,q,\Delta)$ defined in (\ref{FF-dot-Exp}),
that is,
the nontrivial dependence of the polarization vector
\be
\xi^{\mu}(p_1) \thicksim p^{\mu}_1/m_1   \label{polarization-vector-basis}
\ee
on the momentum $p_1$ of the longitude components of vector fields $P^{\ast}$ and $\widetilde{P}^{\ast}$.
In a theory of massive vector bosons $P^{\ast}$ that result from spontaneously broken gauge theories,
the divergence from $\xi^{\mu}(p_1)$ can be automatically canceled in the sum of all diagrams contributing to a given process (known as the Goldstone boson equivalence theorem\cite{Peskin}),
including the $s$ channels and $t$ channels, generated from both the interactions of $P^{\ast} \widetilde{P}^{\ast} \phi$ and the self-interactions of vector bosons $P^{\ast}$ and $\widetilde{P}^{\ast}$,
however,
in the $SU(3)$ ChPT we used here,
the divergence from $\xi^{\mu}(p_1)$ cannot be automatically canceled, because
the massive vector bosons $P^{\ast}$ are only defined as massive matter fields rather than gauge fields of a $SU(4)$ theory, and, only the $t$ channel processes are considered.
So, in addition to the factor $F(k,\Lambda,m_\phi)$,
instead of that
an approximation of $\xi^{\mu}(p_1) \rightarrow 1$  was taken
in Ref. \cite{SchE-scheme-CPC-A-Note},
we will introduce another factor with a power-counting law on $p,q$ for each $P^{\ast} \widetilde{P}^{\ast} \phi$ vertex, as (a Gaussian type one)
\ba
\mathbb{F}(|{\bm p}|,|{\bm q}|,\Delta) = \exp\left[ - \frac{|{\bm p}|^2 + |{\bm q}|^2 }{2 \Delta^2}\right], \label{define-NewExp-factor-2}
\ea
where $p$ ($q$) is the relative momentums of particles in the initial (final) state,
and $\Delta$ is also a typical momentum;
we would treat (\ref{define-NewExp-factor-2}) as the instantaneous approximation form of a covariant factor
$\exp\left[  \frac{p^2  +  q^2 }{2 \Delta^2}\right]$, to ensure the covariant property
and remove the singularity of (\ref{define-NewExp-factor-2}) on the complex $p$- and $q$-plane;
so, equivalently for the full interaction kernel $\overline{K}$, we will introduce a complete modification factor
\ba
\mathcal{F}(k,\Lambda,m_\phi;p,q,\Delta)
&\equiv& [F(k,\Lambda,m_\phi)]^2 \cdot [\mathbb{F}(|{\bm p}|,|{\bm q}|,\Delta)]^2 \non\\
&=& \left[  \frac{\Lambda^2 - m_\phi^2}{\Lambda^2 + {\bm k}^2}  \right]^2
\cdot
\exp\left[ - \frac{|{\bm p}|^2 + |{\bm q}|^2 }{\Delta^2}\right] .  \label{FF-dot-Exp}
\ea
The Gaussian type factor is more reasonable if we recall that a stable BSWF $\chi(p)$ without dispersion in the space should be a Gaussian type, and, in this case, the $\Delta$ should be of the same order of the typical (or, average) value of the relative momentum $p$. To compare with the results in Ref. \cite{SchE-scheme}\cite{SchE-scheme-CPC-A-Note}, the case of $\Delta=+\infty$ will be also considered in our work.

\section{Lorentz Structure of the Bethe-Salpeter Wavefunction}

In this work,
we will treat the orbit angular momentum $L$ and the total spin  $S$ as good quantum numbers, and
we only consider the ground state of a scalar (total angular momentum $J=0$) bound system of $D^{\ast}\bar{D}^{\ast}$ or $B^{\ast}\bar{B}^{\ast}$.
As a ground state, the orbital angular momentum should be $L=0$ and the total spin  should be $S=0$, so that the parity should be $+$ and the C-parity should be $+$, i.e.,
$J^{PC}=0^{++}$.
Neither will we consider the mixing between the states such as $|L=0,S=0 \rangle$, $|L=1,S=1 \rangle$ and $|L=2,S=2 \rangle$.

Generally, in the momentum space, for a bound state system with the total angle-momentum quantum number $J=0$ constituted by a vector particle and its antiparticle, e.g.,  $D^{\ast}\bar{D}^{\ast}$ or $B^{\ast}\bar{B}^{\ast}$, the Lorentz tensor structure (without spinor indices) of the Bethe-Salpeter wavefunction, i.e., the hadronic matrix element defined in
(\ref{projection-of-BS-state},\ref{define-conj-BSWF}), can be expressed with only the metric tensor $g^{\mu\nu}$, the Levi-Civita tensor $\epsilon_{\alpha\beta\gamma\delta}$ and the momentum $p_{1,2}$ of the constituent particles (or equivalently the center-of-mass momentum $P$, and the relative momentum $p$). The polarization vector of a vector particle with momentum $p_1$ in a bound state system , noted as $\xi^{\mu}(p_{1})$, should be ill-defined and not be qualified to express the Lorentz tensor structure, since the momentum $p_1$ is off-shell! So, the Lorentz structure in Ref. \cite{BSE-DstarDstar-ChenXiaozhao-2013}
is not good.

For a bound state $|P\rangle$ of two vector fields $A^{\mu}\widetilde{A}^{\nu}$ with parity eigenvalue $C_P=\pm 1 $, i.e.,
$\hat{P}|P\rangle= C_P |P\rangle $ ,
by recalling $\hat{P}|0\rangle=|0\rangle$ and inserting $\hat{P}^{\dagger}\hat{P}=1$, for the $\mu\nu$ component of a BS wavefunction $\chi^{\mu\nu}$ we can have ($\mu\nu$ only for labelling component rather than for tensor structure)
\ba
\chi^{(I)\mu\nu}_{P}(x_{1},x_{2})&=& \langle \Omega|{\rm T}\{A_{1}^{\mu }(x_{1})\widetilde{A}_{2}^{\nu }(x_{2}) \}|P,\varsigma\rangle \nonumber\\
&=& \langle \Omega|{\rm T}\{\hat{P}^{\dagger}\hat{P}A_{1}^{\mu }(x_{1})\hat{P}^{\dagger}\hat{P}\widetilde{A}_{2}^{\nu }(x_{2})\hat{P}^{\dagger}\hat{P} \}|P,\varsigma\rangle \nonumber\\
&=& \langle \Omega|\hat{P}^{\dagger}{\rm T}\{\hat{P}A_{1}^{\mu }(x_{1})\hat{P}^{\dagger}\hat{P}\widetilde{A}_{2}^{\nu }(x_{2})\hat{P}^{\dagger} \}\hat{P}|P,\varsigma\rangle \nonumber\\
&=& \langle {\Omega}'|  {\rm T}\{ {A'}_{1}^{\mu }({x'}_{1}) {\widetilde{A'}}_{2}^{\nu }({x'}_{2}) \}|P',\varsigma'\rangle
\nonumber\\
&=& \langle {\Omega}|  {\rm T}\{ {A}_{1\mu} ({\widetilde{x}}_{1}) {\widetilde{A}}_{2\nu} ({\widetilde{x}}_{2}) \}\cdot C_P |P,\varsigma\rangle
\nonumber\\
&=&  C_P \cdot  \chi^{(I)\mu\nu}_{P}({\widetilde{x}}_{1},{\widetilde{x}}_{2})  ,\label{P-constraint-x-00}
\ea
where there is $\widetilde{x}=(x^0,-{\bf x})$ and the ``$=$" between a contravariant component and a covariant component holds only in the sense of equal value;
or, (\ref{P-constraint-x-00}) can be written as
\be
\chi^{(I)\mu\nu}_{P}(x_{1},x_{2})=   C_P \cdot  O(\mu,\nu ) \chi^{(I)\mu\nu}_{P} ({\widetilde{x}}_{1},{\widetilde{x}}_{2})  \,,
\label{P-constraint-x}
\ee
where $O(\mu,\nu )$ is a coefficient matrix with the values $O(0,0)=+1$, $O(0,i)=-1$, $O(i,0)=-1$, $O(i,j)=+1$ and $i,j=1,2,3$.
Similarly, in the momentum space we will have
\ba
\chi^{(I)\mu\nu}_{P}(p_{1},p_{2})&=&  C_P \cdot \chi^{(I)\mu\nu}_{P} ({\widetilde{p}}_{1},{\widetilde{p}}_{2})    \,,   \label{P-constraint-1} \\
\chi^{(I)\mu\nu}_{P} (P,p )\,\, &=&   C_P \cdot \chi^{(I)\mu\nu}_{P} ({\widetilde{P}},{\widetilde{p}})   \,.  \label{P-constraint-2}
\ea
Thus, with the constraint of $P$ parity,
the Bethe-Salpeter wave-function for this $J^{PC}=0^{++}$ state should be only expressed as
\ba
\chi^{(I)\mu\nu}_{P} (p)   &=&  f_0(p)g^{\mu\nu}+\frac{1}{M^2}f_1(p)P^{\mu}P^{\nu} \non\\
&&+\frac{1}{M^2}f_2(p)P^{\mu}p^{\nu} +\frac{1}{M^2}f_3(p)p^{\mu}P^{\nu}  + \frac{1}{M^2} f_4(p)p^{\mu}p^{\nu} ,\label{BSWF-P-constraint}
\ea
and
the scalar functions $f_i(p)$ in (\ref{BSWF-P-constraint}) should be constrained to be $f_i(p)=f_i(\widetilde{p})$ and $f_i(p)=f_i(\bar{p})$, with $\widetilde{p}=(p^0,-{\bf p})$ and $\bar{p}=(-p^0,{\bf p})$, so there should be $f_i(p)=f_i(-p)$, that is, $f_i(p)$ is even of $p$.

Similarly, for a bound state $|P\rangle$ with C-parity eigenvalue $C=\pm 1 $, i.e.,
$\hat{C}|P\rangle= C  |P\rangle$ ,
we can have
\be
\chi^{(I)\mu\nu}_{P}(p_{1},p_{2})=  C \cdot  \chi^{(I)\mu\nu}_{P}(p_{2},p_{1})  \,,
\label{C-constraint-1}
\ee
or
\be
\chi^{(I)\mu\nu}_{P}(P,p ) =   C \cdot  \chi^{(I)\mu\nu}_{P}(P,-p)  \,.
\label{C-constraint-2}
\ee
Thus, with the constraint of $C$ parity,
the Bethe-Salpeter wave-function $\chi^{(I)\mu\nu}_{P}$ expressed in (\ref{BSWF-P-constraint}) for this $J^{PC}=0^{++}$ state,
should be invariant under the exchange of $p_{1}\leftrightarrow p_{2}$, or $\chi^{\mu\nu}$ should be even of $p=\frac{1}{2}(p_{1} - p_{2})=\frac{1}{2}(p_{1}^2 + p_{2}^2-2p_{1}\cdot p_{2})$, or, equivalently,  be even of  $p_{1} \cdot p_{2}$ or $P\cdot p $, that is to say, $f_{0,1,4}(p)$  should be even of  $p$, while $f_{2,3}(p)$  should be odd of $p$.

Besides, for convenience in the following sections, we define
\be
V^{\mu}=P^{\mu}/M ,\,p_{l} \equiv  V \cdot p   ,\,p_{t}^{\mu}\equiv p^{\mu}- p_l V^{\mu}  \,\Rightarrow \,p^{\mu}=p_{l} V^{\mu}+p_{t}^{\mu},\label{define-pt}
\ee
where $V$ is the 4-velocity of the bound state, $p_{t}$ is the transverse momentum and $V \cdot p_{t}=0$.

By combining the parity and the C-parity constraints above, we can get
\be
\mbox{$f_{2,3}=0$ and $f_{0,1,4}$ is even of $p$;}
\ee
so the Bethe-Salpeter wave-function for this $J^{PC}=0^{++}$ state should be
\ba
\chi^{(I)\mu\nu}_{P}(p) &=& f_0(p)g^{\mu\nu}+f_1(p)V^{\mu}V^{\nu}  + \frac{1}{M^2} f_4(p)p^{\mu}p^{\nu} \nonumber\\
&=&  f_0(p)g^{\mu\nu} +  f_1(p)V^{\mu}V^{\nu}  + \frac{1}{M^2} f_4(p)\left( p_{t}+ p_l V \right)^{\mu}  \left( p_{t}+ p_l V \right)^{\nu} \nonumber\\
&=&  f_0(p)g^{\mu\nu} +  f_1(p)V^{\mu}V^{\nu}  + \frac{1}{M^2} f_4(p) \left[p_{t}^{\mu}p_{t}^{\nu}
+ p_l (V^{\mu}p_{t}^{\nu}  +  p_{t}^{\mu}V^{\nu})
+ p_l^2 V^{\mu}V^{\nu} \right] \nonumber\\
&=&  f_0(p)g^{\mu\nu}
+[ f_1(p) + \frac{p_l^2}{M^2} f_4(p)] V^{\mu}V^{\nu}
+   \frac{1}{M^2}f_4(p) p_{t}^{\mu}p_{t}^{\nu}\nonumber\\
&&
 +  \frac{p_l}{M^2}f_4(p) (V^{\mu}p_{t}^{\nu}  +  p_{t}^{\mu}V^{\nu}) .   \label{BSWF-J0-A}
\ea

At last, we want to point out that,
different gauge fixed conditions for the vector fields will lead to different constraints on the BS wavefunction $\chi^{(I)\alpha\mu}_{P}$, although the physical transition amplitudes should be independent on the choice of gauge fixed conditions. For example, in the
Lorentz gauge, there will be automatically $p_{1\alpha} \chi^{(I)\alpha\mu }_{P} =p_{2\mu} \chi^{(I)\alpha\mu }_{P} =0$ by inserting $p_{1\alpha}$ or $p_{2\mu}$ into the propagator $S_1^{\alpha\sigma}(p_1)$ or  $S_2^{\mu\gamma}(p_2)$ in the r.h.s. in BS equation, however, in the unitary gauge, there will be $p_{1\alpha} \chi^{(I)\alpha\mu }_{P} , p_{2\mu} \chi^{(I)\alpha\mu}_{P}  \neq 0$ since $p_{1,2}$ are off-shell! As we have chosen the interaction Lagrangian terms listed in (\ref{L-P-P-V}-\ref{L-Pstar-P-V}), that is to say, we have chosen the unitary gauge for the field $P^*$ and their antiparticles $\widetilde{P}^*$, we cannot use the Lorentz gauge condition to constraint $p_{1\alpha} \chi^{(I)\alpha\mu }_{P} =p_{2\mu} \chi^{(I)\alpha\mu }_{P} =0$  any more! So the form of the BSWF in Ref. \cite{BSE-DstarDstar-ChenXiaozhao-2013},\cite{BSE-DstarDstar-ChenXiaozhao-2015} are wrong.

\section{Solve the BSE in the Rest Frame}

\subsection{Solve BSE in the rest frame and the covariant instantaneous approximation}

As in (\ref{define-pt},\ref{define-k}), we define $q_{l} \equiv  V \cdot q $, $q_{t}^{\mu}\equiv q^{\mu}- q_l V^{\mu}$, then we will have  $q^{\mu}=q_{l} V^{\mu}+q_{t}^{\mu}$, $q^2 =  q_{l}^2  +  q_{t}^{2}$, and $k= p_1-q_1 = p-q $ and $k_l= p_l-q_l ,\, k_t= p_t-q_t $.
The results of the scalar functions $f_i(p)$ are independent on the reference frame,
so, for convenience,
we will solve the BSE in the rest frame, i.e., there will be
\ba
&&P=(M,0),\, V=(1, {\bm 0});\\
&&p_l = p^0,\, q_l = q^0,\, p_{t}^{\mu}=p^{\mu}- p_{l}V^{\mu}= (0,{\bm p}).\label{pl=p0}
\ea

By inserting the BS wavefunction (\ref{BSWF-J0-A}) and the propagators (\ref{propagator}) of $P^{\ast}$ and $\widetilde{P}^{\ast}$ into the BS equation (\ref{BSE-isospin}), we can get the equation below:
\ba
\chi^{(I) \alpha\mu}_{P }(p)  &=&  f_0(p)g^{\alpha\mu}
+[ f_1(p) + \frac{p_l^2}{M^2} f_4(p)] V^{\alpha}V^{\mu}
+   \frac{1}{M^2}f_4(p)   \frac{1}{1+\frac{|{\bm p}|^2}{M^2}}          p_{t}^{\alpha}p_{t}^{\mu}\nonumber\\
&&
 +  \frac{p_l}{M^2}f_4(p) (V^{\alpha}p_{t}^{\mu}  +  p_{t}^{\alpha}V^{\mu})
 \nonumber\\
&=&   \frac{-i(g^{\alpha\sigma}-\frac{p_1^{\alpha}p_1^{\sigma}}{m_1^2})}{p_1^2-m_1^2}
\cdot \frac{-i(g^{\mu\gamma}   -\frac{p_2^{\mu}p_2^{\gamma}}{m_2^2})}   {p_2^2-m_2^2}
\cdot \int {d^4 q  \over (2\pi)^{4}}
{\ovl K}^{total} _{\sigma\gamma\nu\beta}(P,p,q)
\nonumber\\
&&
\cdot  [ f_0(q)g^{\nu\beta}
+[ f_1(q) + \frac{q_l^2}{M^2} f_4(q)] V^{\nu}V^{\beta}
+   \frac{1}{M^2}f_4(q)      \frac{1}{1+\frac{|{\bm q}|^2}{M^2}}         q_{t}^{\nu}q_{t}^{\beta}   \nonumber\\
&&
 +  \frac{q_l}{M^2}f_4(q) (V^{\nu}q_{t}^{\beta}  +  q_{t}^{\nu}V^{\beta}) ]  ,  \nonumber\\
 \label{BSE-pq-J0I0-d4q}
\ea

In the system of heavy-flavor hadrons, if the typical scale for the exchanged momentum  $k$ is much less than the mass of heavy-flavor hadrons,
for instance,  $k$ at the same order of QCD typical scale $\Lambda_{QCD}$, then
it will be reasonable and convenient to take
the so-called
covariant instantaneous approximation (C.I.A) $  {\ovl K}^{ij,i'j'} (P,p,q) =  {\ovl K}^{ij,i'j'} (P, p_t, q_t)$  in the calculations, i.e.,
by taking
\ba
p_l = q_l   \quad\mbox{only in ${\ovl K}$}\quad\mbox{(C.I.A)}
\ea
in the interaction kernel, where there is $p_l = q_l =0$ in a system of particle pair.
More strictly to say, in the numerator of the propagator of $P^{\ast}$ and $\widetilde{P}^{\ast}$,
there should also be  $p_l = q_l =0$ after taking the covariant instantaneous approximation, since
the gauge fixed condition uniquely corresponding to the numerator of the propagator
is essentially a kind of interaction effect, partly to determine the full interaction kernel.
Particularly, in the rest frame (R.F.) case , due to $p^0=p^l$, $q^0=q^l$, the covariant instantaneous approximation is equivalent to
\ba
p^0 = q^0 =0   \quad\mbox{only in ${\ovl K}$}\quad\mbox{(C.I.A \&  R.F. )},
\ea
which is
just the so-called on-shell approximation.

If we define
\ba
\Phi_1(p_t^2) &\equiv&  \int \frac{dp_l}{2 \pi} f_0(p), \\
\Phi_2(p_t^2) &\equiv&  \int \frac{dp_l}{2 \pi} [ f_1(p) + \frac{p_l^2}{M^2} f_4(p)],\\
\Phi_3(p_t^2) &\equiv&  \int \frac{dp_l}{2 \pi} f_4(p),\\
\Phi_4(p_t^2) &\equiv&  \int \frac{dp_l}{2 \pi}  \frac{p_l}{M^2}f_4(p),
\ea
we will have $\Phi_4(p_t^2) =0$ since $f_{0,1,4}(p^0,{\bm p})$ are even of $p^0$ or $p_l$; furthermore, we can define
\ba
\bar{\Phi}_{P}^{(I)\alpha\mu }( {\bm p} )
\equiv \int {d p^0 \over (2\pi) }
 \bar{\chi}_{P}^{(I) \alpha\mu }(p)
&=&  \Phi_1(p_t^2)  g^{\alpha\mu}
+\Phi_2(p_t^2) \frac{ P^{\alpha}P^{\mu}}{M^2}
+\Phi_3(p_t^2)      \frac{1}{1+\frac{|{\bm p}|^2}{M^2}}       \frac{ p_{t}^{\alpha}p_{t}^{\mu} }{M^2} , \non\\
\\
\Phi_P^{(I) \nu\beta}(  {\bm q} )
\equiv\int {d q^0  \over (2\pi) }
 \chi_P^{(I)\nu\beta}(q)
&=&
 \Phi_1(q_t^2) g^{\nu\beta}
+\Phi_2(q_t^2)  \frac{P^{\nu}P^{\beta} }{M^2}
+\Phi_3(q_t^2)     \frac{1}{1+\frac{|{\bm q}|^2}{M^2}}     \frac{  q_{t}^{\nu}q_{t}^{\beta} }{M^2}     .\non\\
\label{SchWF-VS-BSWF}
\ea
In the case of $p^l=p^0$, as shown in (\ref{pl=p0}), the functions $\Phi_P^{(I) \nu\beta}(  {\bm q} )$ are indeed the Schrodinger wavefunction\cite{Lurie}\cite{BSWF-SchWF}.
Then,
in the rest frame of the bound state system, after performing the integration $\int dp_l$ or $\int dp^0$, in both sides of (\ref{BSE-pq-J0I0-d4q}) and taking the covariant instantaneous approximation, we can get
\ba
\Phi_{P}^{(I)\alpha\mu }(  p )
&=& \int {d^3 q  \over (2\pi)^{3}}
\left[\int dp^0
\frac{1}{p_1^2-m_1^2}
\cdot \frac{1}{p_2^2-m_2^2}  \right]\nonumber\\
&&\cdot
[-i(g^{\alpha\sigma}-\frac{p_1^{\alpha}p_1^{\sigma}}{m_1^2})]
[-i(g^{\mu\gamma}   -\frac{p_2^{\mu}p_2^{\gamma}}{m_2^2})]
{\ovl K}^{total} _{\sigma\gamma\nu\beta}(P,p,q)
\cdot    \Phi_P^{(I) \nu\beta}(  q )
\nonumber\\
&=& \int {d^3 q  \over (2\pi)^{3}}   C_{E}    \mathcal{K}^{\alpha\mu}_{\nu\beta}(P,p,q) \Phi_P^{(I) \nu\beta}(  q )   , \label{BSE-pq-J0I0-d3q}
\ea
where there is
\ba
C_{E}   &\equiv&\int_{-\infty}^\infty\,dp^0
{1\over (p_1^2 - m_1^2 + i\epsilon)}{1\over(p_2^2 - m_2^2+ i\epsilon)} \\
&=& -i\pi\, { (E_1+E_2)/E_1E_2\over E^2-(E_1+E_2)^2}  ,\label{define-CE}
\ea
with $E_i \equiv \sqrt{{\bf p}^2 + m_i^2}$, $E=P^0$,
and
\ba
 \mathcal{K}^{\alpha\mu}_{\nu\beta}(P,p,q) \equiv
 [-i(g^{\alpha\sigma}-\frac{p_1^{\alpha}p_1^{\sigma}}{m_1^2})]
[-i(g^{\mu\gamma}   -\frac{p_2^{\mu}p_2^{\gamma}}{m_2^2})]
{\ovl K}^{total} _{\sigma\gamma\nu\beta}(P,p,q)
.\label{define-Ktotal}
\ea
For the $I=0$ case there is the reduction
\ba
{\ovl K}^{total}_{\sigma\gamma\nu\beta}
&\rightarrow&\left[  {\ovl K}^{11,11} _{\sigma\gamma\nu\beta}(P,p,q)
+ {\ovl K}^{11,22}_{\sigma\gamma\nu\beta}(P,p,q)       \right]\non\\
&=&\left[  3 {\ovl K}^{(\pi)11,11}
+ {\ovl K}^{(\eta)11,11}
+ 3 {\ovl K}^{(\rho)11,11}
+ {\ovl K}^{(\omega)11,11}
+ {\ovl K}^{(\sigma)11,11} \right] _{\sigma\gamma\nu\beta}  \non\\
&=& \sum_{\phi=\pi,\eta,\rho,\omega,\sigma} c^{I}_{\phi} \cdot  {\ovl K}^{(\phi)11,11} _{\sigma\gamma\nu\beta}    ,     \label{kernel-J0I0}
\ea
with
$c^{I}_{\pi}=3,c^{I}_{\eta}=1,c^{I}_{\rho}=3,c^{I}_{\omega}=1,c^{I}_{\sigma}=1$,
while for the $I=1$ case there is a similar reduction with
$c^{I}_{\pi}=-1,c^{I}_{\eta}=1,c^{I}_{\rho}=-1,c^{I}_{\omega}=1,c^{I}_{\sigma}=1$.

Furthermore, with the expressions of propagators in
(\ref{propagator},\ref{propagator-00}) we can rewrite the BSE (\ref{BSE-isospin}) as
\ba
\chi^{(I)}_{P\alpha\mu}(p)&=& \mathfrak{S}_{1}  (p_1,m_1)    \mathfrak{S}_{2}  (p_2,m_2)\non\\
&& \cdot \mathbb{S}_{1 \alpha\sigma}  (p_1,m_1) \mathbb{S}_{2  \mu\gamma}  (p_2,m_2) \cdot \int d^4q \overline{K}^{(total)\sigma\gamma\nu\beta}(P,p,q)
\chi^{(I)}_{P \nu\beta}(q),\non\\
\\
\mbox{or}\non\\
\int dp^0 \chi^{(I)}_{P\alpha\mu}(p)&=& \left[ \int dp^0  \mathfrak{S}_{1}  (p_1,m_1) \mathfrak{S}_{2}  (p_2,m_2)\right]    \non\\
&& \cdot \mathbb{S}_{1 \alpha\sigma}  (p_1,m_1) \mathbb{S}_{2 \mu\gamma}  (p_2,m_2) \cdot \int d^4q \overline{K}^{(total)\sigma\gamma\nu\beta}(P,p,q) \chi^{(I)}_{P\nu\beta}(q),\non\\
\\
\widetilde{\chi}^{(I)}_{P\alpha\mu}(p)&=&
C_{E}
\cdot\mathbb{S}_{1 \alpha\sigma}  (p_1,m_1) \mathbb{S}_{2 \mu\gamma}  (p_2,m_2) \cdot \int d^4q \overline{K}^{(total)\sigma\gamma\nu\beta}(P,p,q) \chi^{(I)}_{P\nu\beta}(q),\non\\
\\
\left[C_{E}\right]^{-1}  \widetilde{\chi}^{(I)}_{P\alpha\mu}(p)&=&
\mathbb{S}_{1 \alpha\sigma}  (p_1,m_1) \mathbb{S}_{2  \mu\gamma}  (p_2,m_2) \cdot \int d^4q \overline{K}^{(total)\sigma\gamma\nu\beta}(P,p,q) \chi^{(I)}_{P\nu\beta}(q),\non\\
\\
\chi^{(I)}_{P\alpha\mu}(p)&=& \mathfrak{S}_{1}  (p_1,m_1)    \mathfrak{S}_{2}  (p_2,m_2)
\cdot
\left[C_{E}\right]^{-1}  \widetilde{\chi}^{(I)}_{P\alpha\mu}(p), \label{BSWF0123=SSdot123}
\ea
that is to say, we can get
\ba
\chi^{(I)}_{P \alpha\mu}(p) =  {1\over (p_1^2 - m_1^2 + i\epsilon)}{1\over(p_2^2 - m_2^2+ i\epsilon)}
\cdot
\left[C_{E}\right]^{-1}  \widetilde{\chi}^{(I)}_{P \alpha\mu}(p). \label{BSWF0123=SSdot123}
\ea

\subsection{Normalization of BS wavefunctions}

Firstly, according to (\ref{define-I-tensor}), we have
\ba
&&I_{\sigma\gamma\nu\beta}(P,p,q) =    I_{\sigma\gamma\nu\beta}(p_1,p_2,k=p-q) \non\\
&\equiv&
(2\pi)^4 \delta^{4}(q-p) \cdot
S_{1 \nu\sigma}^{-1} (p_1,m_1)   S_{2 \beta\gamma}^{-1} (p_2,m_2)   \non\\
&= &
(2\pi)^4 \delta^{4}(q-p) \cdot
\left[ \frac{-i(g^{ \nu\sigma}-\frac{p_1^{\nu}p_1^{\sigma}}{m_1^2})}{p_1^2-m_1^2} \right]^{-1}
\cdot
\left[ \frac{-i(g^{\beta\gamma}   -\frac{p_2^{\beta}p_2^{\gamma}}{m_2^2})}{p_2^2-m_2^2} \right]^{-1} \non\\
&= &
(2\pi)^4 \delta^{4}(q-p) \cdot
\left[i (p_1^2-m_1^2) \cdot        \frac{(g_{ \nu\sigma}-\frac{p_{1\nu}p_{1\sigma}}{m_1^2})}
{(g_{ \nu'\sigma'}-\frac{p_{1\nu'}p_{1\sigma'}}{m_1^2})(g^{ \nu'\sigma'}-\frac{p_1^{\nu'}p_1^{\sigma'}}{m_1^2})}  \right] \non\\
&&
\cdot
\left[i (p_2^2-m_2^2) \cdot        \frac{(g_{\beta\gamma}   -\frac{p_{2\beta}p_{2\gamma}}{m_2^2})}
{(g_{\beta'\gamma'}   -\frac{p_{2\beta'}p_{2\gamma'}}{m_2^2})(g^{\beta'\gamma'}   -\frac{p_2^{\beta'}p_2^{\gamma'}}{m_2^2})} \right]   ,  \non\\
\ea
then we can compute $\frac{\partial I}{\partial P^0}$ with
\ba
p_{1}^{\mu} &=& \eta_1 P^{\mu} + p^{\mu} = \eta_1 P^{\mu}+p_t^{\mu},\\
p_2^{\mu} &=& \eta_2 P^{\mu} - p^{\mu} = \eta_2 P^{\mu}-p_t^{\mu},\\
k^{\mu} &=& p_{1}^{\mu} - q_1^{\mu} = p_t^{\mu}-q_t^{\mu}.\label{momentum-relations-ptqt}
\ea

Secondly, in the ladder approximation to solve BSE, only t-channels are considered and the factor $\frac{1}{P^2-m_\phi^2}$ in s-channels are ignored,
so, if the vertices and the propagators of exchanged mesons is independent on $P$, then we will have $\frac{\partial \overline{K}^{total}}{\partial P^{0}} =0$.
However, in the ChPT, the vertices are dependent on $P$ and we will have
$\frac{\partial \overline{K}^{total}}{\partial P^{0}} \neq 0$!
With $\frac{\partial k}{\partial P^0} =0$, see (\ref{momentum-relations-ptqt}), we indeed have
\ba
&&\frac{\partial}{\partial P^0}{\ovl K}^{total}_{\sigma\gamma\nu\beta}(P,p,q)
=\frac{\partial}{\partial P^0}{\ovl K}^{total}_{\sigma\gamma\nu\beta}(p_1,p_2,k ) \non\\
&=& \frac{\partial}{\partial p_1}{\ovl K}^{total}_{\sigma\gamma\nu\beta}(p_1,p_2,k) \cdot\frac{\partial p_1}{\partial P^0} +
 \frac{\partial}{\partial p_2}{\ovl K}^{total}_{\sigma\gamma\nu\beta}(p_1,p_2,k) \cdot\frac{\partial p_2}{\partial P^0}     .
\ea

Thirdly, after inserting BS wave-functions expressed in (\ref{BSWF0123=SSdot123}), we can perform the two integrations,
\ba
R_{I} \equiv i  \int {d^4q d^4p \over (2\pi)^{8}}
\frac{1}{\sqrt{2}}\bar{\chi}_{P}^{(I)\sigma\gamma }(p)
\cdot
\frac{\partial}{\partial P^0}\left[
I_{\sigma\gamma\nu\beta}(P,p,q)
  \right]  \cdot
 {1 \over 2E_{\bf P}} \cdot
\frac{1}{\sqrt{2}} \chiP^{(I)\nu\beta}(q) \,,\label{BSWF-normal-condition-pq-RI}
\ea
and
\ba
R_{K} \equiv i  \int {d^4q d^4p \over (2\pi)^{8}}
\frac{1}{\sqrt{2}}\bar{\chi}_{P}^{(I)\sigma\gamma }(p)
\cdot
\frac{\partial}{\partial P^0}\left[
{\ovl K}_{\sigma\gamma\nu\beta}(P,p,q)
  \right]  \cdot
 {1 \over 2E_{\bf P}} \cdot
\frac{1}{\sqrt{2}} \chiP^{(I)\nu\beta}(q) ,\label{BSWF-normal-condition-pq-RK}
\ea
by integrating out $p^0$ and $q^0$ with the residue.
Furthermore, for the consistency,
the covariant instantaneous approximation now still holds in (\ref{BSWF-normal-condition-pq-RK}), so, there is still $p^0=q^0=0$ in the kernel $ {\ovl K}^{total}$,
which means, $ {\ovl K}^{total}$ is independent on $p^0$ and $q^0$;
besides, the factor $[F(k,\Lambda,m_\phi)]^2  \cdot \mathbb{F}(|{\bm p}|,|{\bm q}|)$ in  (\ref{FF-dot-Exp}) is also independent on both $p^0$, $q^0$  and  $P^0$.
So, we can have
\be
R_K  =  i  \int { d^3{\bm p} d^3{\bm q}  \over (2\pi)^{6} }
\bar{\Phi}_{P}^{\sigma\gamma }(  p )
\cdot
\frac{\partial}{\partial P^0}\left[
  {\ovl K}_{\sigma\gamma\nu\beta}(P,p,q)\right]_{p^0=q^0=0}  \cdot
 {1 \over 2E_{\bf P}}\,
\Phi_P^{\nu\beta}(  q ) ,\,,\label{BSWF-normal-condition-pq-RK}
\ee
which is only of a 3-momentum integration.
At last, by combining (\ref{BSWF-normal-condition-pq-RI}) and (\ref{BSWF-normal-condition-pq-RK}), the normal condition (\ref{BSE-norm-isospin}) can be expressed as
\ba
R_I-R_K =   1   \,,\,  ( P^0=E_{\bf P}) \,.
\ea
The numerical results in later sections shows that the values of $R_{I}$ and $(-R_K )$ are about at the same order. \\

\newpage
\section{Decay Width of the Molecular States}

\begin{figure}[!htbp]
\centering \includegraphics[scale=0.6]{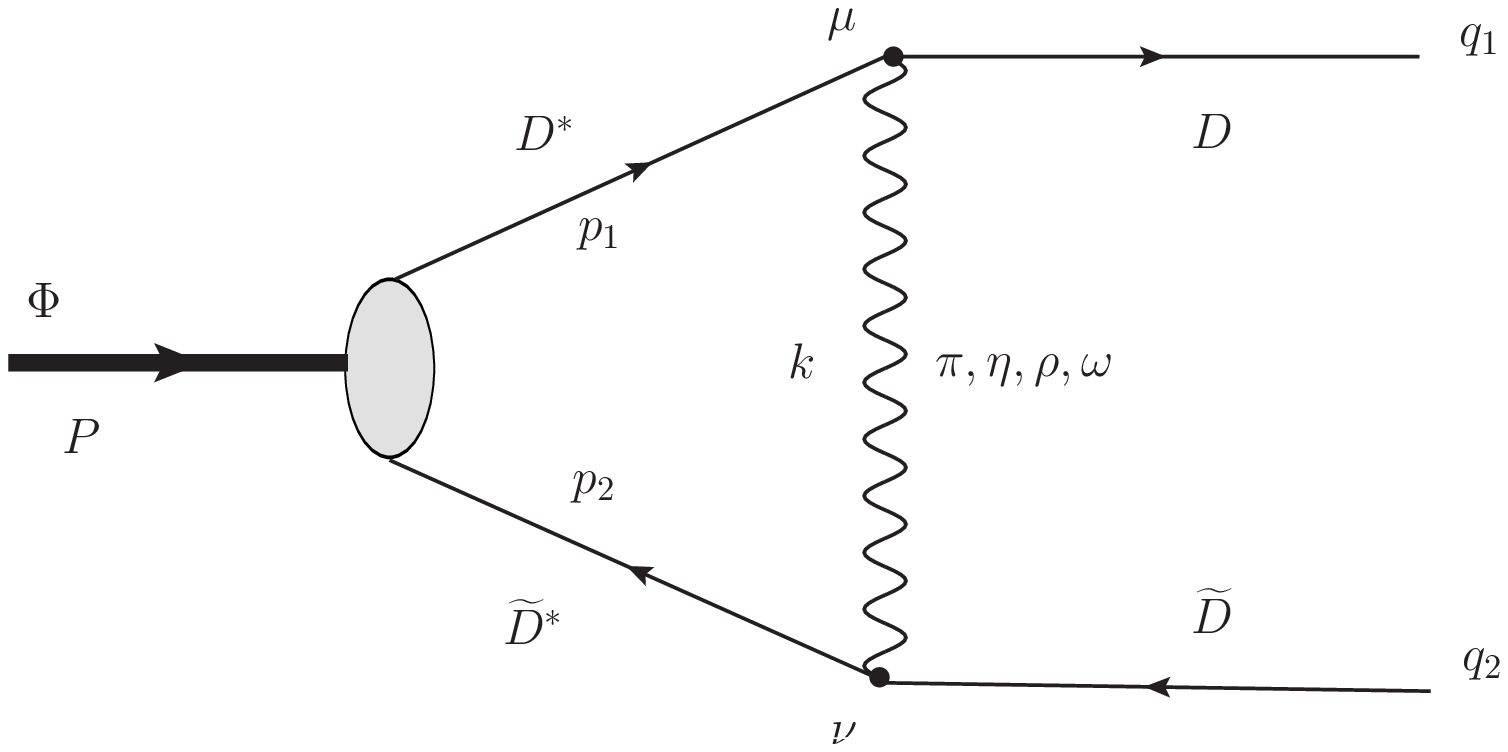} \\
\centering  (a)\\
\centering \includegraphics[scale=0.6]{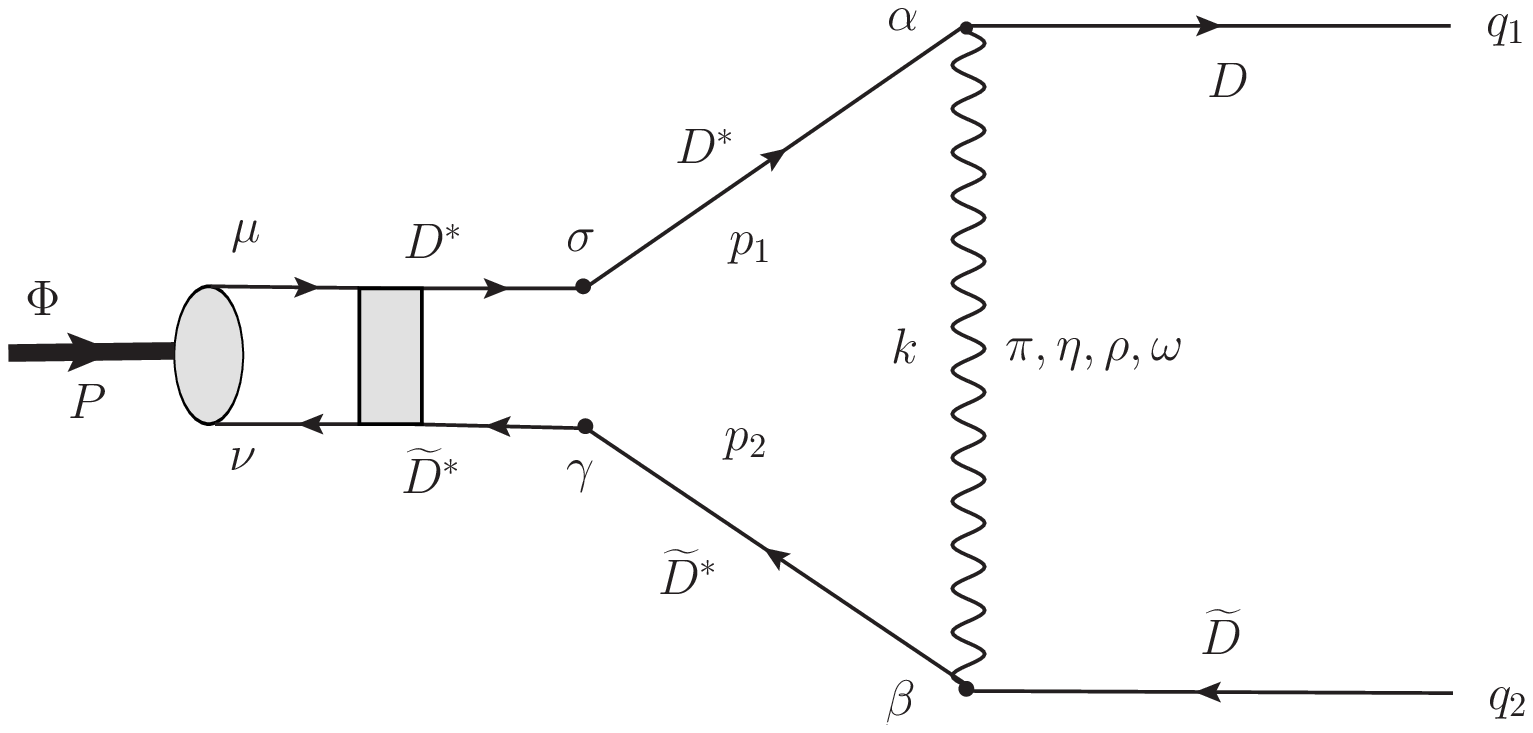} \\
\centering  (b)\\
\centering \includegraphics[scale=0.6]{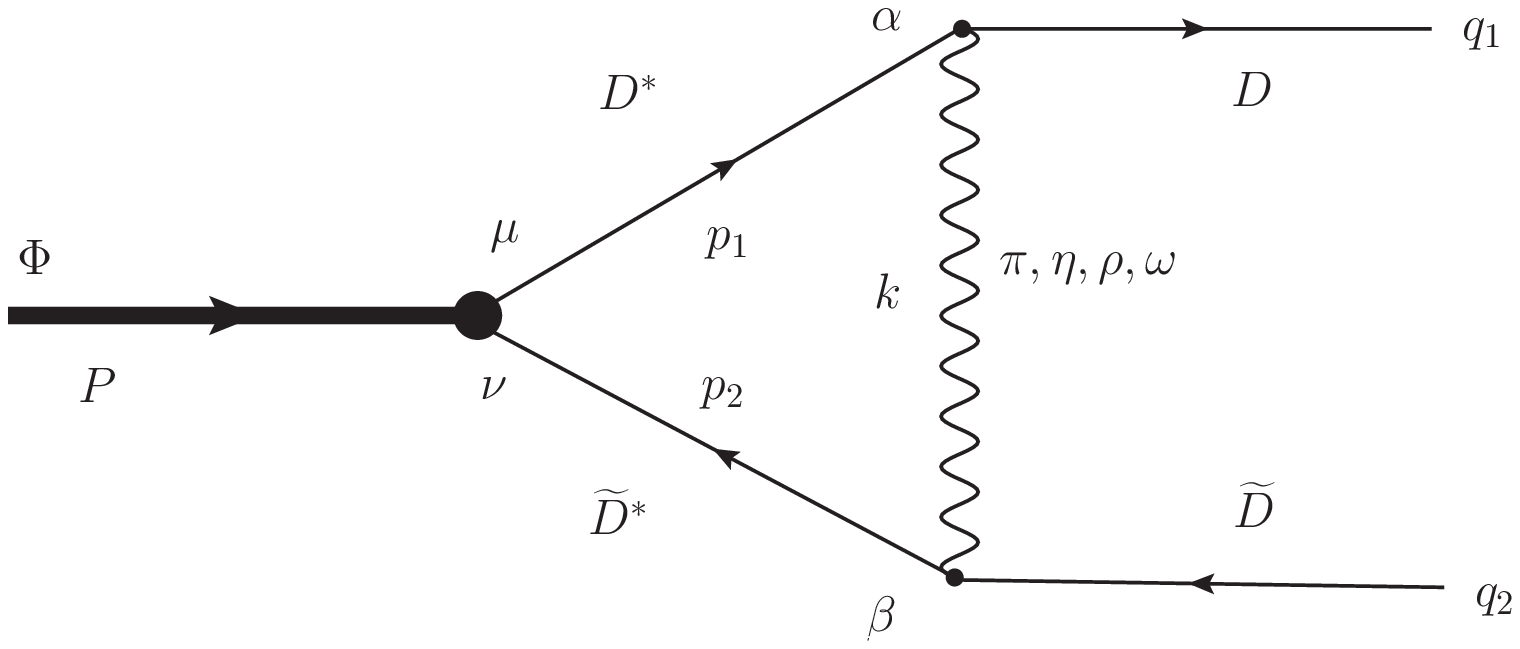} \\
\centering  (c)\\
\caption{
\label{Decay-Dstar-Dstarbar}
(a) Feynman diagram of the decay process $\Phi\rightarrow D^{+} D^{-}$ with a gray box or circle as an effective vertex, where $\Phi$ is the bound state of $D^{\ast} \widetilde{D}^{\ast}$, and the contributions from the $s$-channel processes are omitted since it will be suppressed; (b) is equal to (a) according to the BSE; (c) is the Feynman diagram of $\Phi\rightarrow D^{+} D^{-}$ by treating $\Phi$ as an elementary particle with an effective coupling to the $D^{\ast}$ and $\widetilde{D}^{\ast}$ mesons.
}
\end{figure}

Let's now consider the decay process $\Phi\rightarrow D^{+} D^{-}$,
 where $\Phi$ is the hadronic molecular state constituted by $D^{\ast}$ meson and its anti-partner $\widetilde{D}^{\ast}$. The Feynman diagram of this process has been shown in Fig. \ref{Decay-Dstar-Dstarbar}-(a), where the contributions from the $s$-channel processes mediated by the light mesons are omitted since it will be suppressed by the large momentum of the light mesons.

The $\mathcal{S}$ matrix in the interaction picture can be written as
(by ignoring the possible mixing between $D^{(\ast) 0}$ and $\bar{D}^{(\ast) 0}$) \cite{WuXingHua-KK}
\ba
\mathcal{S}_{\Phi\rightarrow D^{+} D^{-}}
&\equiv&    1+ i\mathcal{T}_{\Phi\rightarrow D^{+} D^{-}}  \non\\
&=& \sum_{I',I'_3}  \langle D^{+}(q_1),  D^{-}(q_2)|
I',I'_3\rangle \cdot  \langle I',I'_3|   {\rm T}\, e^{i\int d^{4}x \mathcal{L}^{P^*P \phi}_{int}(x) } |\Phi(P);I,I_3\rangle,\non\\
&\equiv& \sum_{I',I'_3}  C_{(I',I'_3)}^{+-}\cdot
\langle f|   {\rm T}\, e^{i\int d^{4}x \mathcal{L}^{P^*P \phi}_{int}(x) } |i\rangle,
\label{decay-amp-0}
\ea
where $\phi$ denotes the light mesons $(\pi,\eta,\rho,\omega,\sigma)$,
$P,q_1,q_2$ are the momentums of fields $\Phi(Z),D(y_1),\widetilde{D}(y_2)$, respectively; $I,I_3$ are the isospin quantum numbers of the initial state;
and, $\mathcal{L}^{P^*P\phi}_{int}$ can be got from (\ref{L-Pstar-P-Pion},\ref{L-Pstar-P-V}), as
\ba
\mathcal{L}^{P^*P\phi}_{int}  &=&
\mathcal{L}_{P^*P\mathbb{P}}       +\mathcal{L}_{\widetilde{P}^*\widetilde{P}\mathbb{P}}
+ \mathcal{L}_{P^{*}P  \mathbb{V}} + \mathcal{L}_{\widetilde{P}^{*}\widetilde{P}  \mathbb{V}}.
\ea

We should keep in mind that $i\mathcal{T}_{\Phi\rightarrow D^{+} D^{-}} $ is indeed an $S$-wave partial amplitude,
so,
the system is symmetric under rotation around the $z$-axis; and,
after defining the spherical angles $(\theta,\varphi)$ with $\theta$ the angle between the direction of $q_1$ and the $z$-axis,
we can take the result of $\varphi=0$ as the results of all $\varphi$ values.
According to (\ref{hc-BSWF},\ref{trivial-BSWF}), by omitting terms of $\mathcal{O}((\mathcal{L}_{int})^4 )$, the leading nonzero terms in (\ref{T-matrix-0}) will be
\ba
&& i\mathcal{T}_{\Phi\rightarrow D^{+} D^{-}} \non\\
&=& {\Big\{}
\sum_{I',I'_3}  C_{(I',I'_3)}^{+-}\cdot  \langle f|
{\rm T}\,\{
\frac{i^{2}}{2!}\int d^{4}x_1 d^{4}x_2
\left[      \mathcal{L}_{P^*P\mathbb{P}}(x_1)   \cdot \mathcal{L}_{\widetilde{P}^*\widetilde{P}\mathbb{P}}(x_2)
+   \mathcal{L}_{P^{*}P  \mathbb{V}}(x_1)  \cdot   \mathcal{L}_{\widetilde{P}^{*}\widetilde{P}  \mathbb{V}}(x_2) \right]
\}|i\rangle
{\Big\}}\non\\
&&
+ \left\{x_1\leftrightarrow x_2\right\}\non\\
&=& {\Big\{}
\sum_{I',I'_3}  C_{(I',I'_3)}^{+-}\cdot  \langle f|
{\rm T}\, \{
\frac{i^{2}}{2!}\int d^{4}x_1 d^{4}x_2
\left[  -  c^2
\cdot P_b  P^{*\dag}_{a\lambda}   \partial^\lambda \mathbb{P}_{ba}(x_1)
\cdot
\tilde{P}^\dag_{a}   \tilde{P}^*_{b\tau}   \partial^\tau\mathbb{P}_{ab}(x_2)
\right]
\}
|i\rangle
+\left[ ... \right]
{\Big\}}\non\\
&& + \left\{x_1\leftrightarrow x_2 \right\} \non\\
&=& {\Big\{}
{\sum_\Psi}
\sum_{I',I'_3}  C_{(I',I'_3)}^{+-}\cdot  \langle f|
{\rm T}\, \{
\frac{i^{2}}{2!}\int d^{4}x_1 d^{4}x_2
\left[  -  c^2
P_b(x_1)  \partial^\lambda \mathbb{P}_{ba}(x_1)
\cdot
\tilde{P}^\dag_{a}(x_2)   \partial^\tau\mathbb{P}_{ab}(x_2)
\right]
\}   |\Psi\rangle\non\\
&&
\cdot\langle\Psi|
{\rm T}\,\{P^{*\dag}_{a\lambda} (x_1)\tilde{P}^*_{b\tau}(x_2)   \}
|i\rangle
+\left[... \right]
{\Big\}}
+ \left\{x_1\leftrightarrow x_2\right\}, \label{T-matrix-0}
\ea
where we define $c \equiv \left(\frac{2g_\pi\sqrt{m_P m_{P^*}}}{f_\pi} \right)$, the ellipsis ``$...$" denotes other terms in   $\mathcal{L}_{P^*P\mathbb{P}}(x_1)   \cdot \mathcal{L}_{\widetilde{P}^*\widetilde{P}\mathbb{P}}(x_2)
+   \mathcal{L}_{P^{*}P  \mathbb{V}}(x_1)  \cdot   \mathcal{L}_{\widetilde{P}^{*}\widetilde{P}  \mathbb{V}}(x_2)$, and
$\{|\Psi\rangle\}$ are a set of complete basis and at the leading order we only consider contributions from the vacuum state $|\Omega\rangle $.

From the relations of BS wave-functions in (\ref{projection-of-BS-state}), we can have
\ba
&&\langle \Omega| {\rm T}\, \{ P^{*\mu}_i (x_1)  \widetilde{P}^{*\nu}_j (x_2) \}    |i\rangle
=   C_{(I,I_3)}^{ij} \cdot  \chi^{(I)\mu\nu}_{P}(x_1,x_2) ,\\
&&\langle \Omega| {\rm T}\, \{ \partial^{\lambda} P^{*\mu}_i (x_1)  \widetilde{P}^{*\nu}_j (x_2) \}   |i\rangle
=   C_{(I,I_3)}^{ij} \cdot (- i p^{\lambda}_1) \cdot \chi^{(I)\mu\nu}_{P}(x_1,x_2) ,\\
&&\langle \Omega| {\rm T}\, \{ P^{*\mu}_i  (x_1) \partial^{\beta} \widetilde{P}^{*\nu}_j(x_2) \} |i\rangle
=  C_{(I,I_3)}^{ij}  \cdot  (-i p^{\beta}_2)    \cdot \chi^{(I)\mu\nu}_{P}(x_1,x_2),\\
&&\langle \Omega| {\rm T}\, \{ \partial^{\lambda} P^{*\mu}_i(x_1)   \partial^{\beta} \widetilde{P}^{*\nu}_j(x_2) \}|i\rangle
= C_{(I,I_3)}^{ij}  \cdot  (- i p^{\lambda}_1)  (-i p^{\beta}_2) \cdot \chi^{(I)\mu\nu}_{P}(x_1,x_2),
\ea
then, after transforming into the momentum space, from the Wick theorem we can get
\ba
i\mathcal{T}_{\Phi\rightarrow D^{+} D^{-}}  &\equiv& (2 \pi)^4 \delta^{(4)}(P-q_1-q_2)\cdot
i\mathcal{M}_{\Phi\rightarrow D^{+} D^{-}} ,\label{decay-amp-define-iM}\\
i\mathcal{M}_{\Phi\rightarrow D^{+} D^{-}} (P,q)
&=&
\sum_{I',I'_3}   \sum_{i'j'}\sum_{ij}
C_{(I',I'_3)}^{+-} C_{(I',I'_3)}^{i'j'}    C_{(I,I_3)}^{ij}   \cdot \non\\
&&
\int \frac{d^4 p  }{(2 \pi)^4}
\chi^{(I)\mu\nu}_{P}(p)
\cdot
K_{\mu\nu}^{ ij,i'j'}(P,p,q)
, \label{decay-amp-iM-form}
\ea
with
\ba
K_{\mu\nu}^{ ij,i'j'}= \sum_{\phi_i=\pi,\eta,\rho,\omega} K_{\mu\nu}^{(\phi_i) ij,i'j'} ,
\ea
where $K_{\mu\nu}^{(\pi,\eta,\rho,\omega) ij,i'j'}$ are the contributions from $\pi,\eta,\rho,\omega$, respectively;
since
the $(P^{\ast} i\overleftrightarrow{\partial} P)\sigma$ type interaction terms are forbidden
by the $P$ partity conservation,
there will be
\ba
K^{(\sigma)  ij,i'j'}_{\alpha\beta}  = 0.
\ea
Due to the isospin  conservation in a strong interaction process, ,
we need only consider the $I'=I,\,I'_3=I_3$ case, that is to say, the sum $\sum_{I',I'_3}$ can be reduced.
The explicit expressions of the effective kernels $K_{\mu\nu}^{(\phi_i) ij,i'j'}$ are listed in Appendix \ref{appendix-kernel-in-decay}.

The covariant instantaneous approximation will not be taken in the decay process, i.e., there would be $p^0, q^0 \neq 0$ in the kernel $ {\ovl K}^{total}$  of $D^{\ast} \widetilde{D}^{\ast} \rightarrow D \widetilde{D} $ processes, although this approximation has been taken in solving the BSE for the kernel of $D^{\ast} \widetilde{D}^{\ast} \rightarrow D^{\ast} \widetilde{D}^{\ast}$ processes.

At last,
we can get the partial decay width
\ba
\Gamma_{\Phi\rightarrow D^{+} D^{-}}  =  \frac{1}{2 M}
\int d \cos \theta  \frac{1}{16 \pi } \frac{2 |{\bm q}_1|}{E_{cm}}     \cdot
\left|
i \mathcal{M}_{\Phi\rightarrow D^{+} D^{-}}(P,q)
\right|^2  ,
\ea
where we will have $E_{cm}= M$ in the rest frame of the initial state $|\Phi\rangle$, with $M$ for the mass and $P$ for the momentum of $\Phi$.
Since $i \mathcal{M}_{\Phi\rightarrow D^{+} D^{-}}(P,q)$ is indeed a partial-wave amplitude, including an integration over the angles, it is independent on the $\theta$ angle, and we can have
\ba
\Gamma_{\Phi\rightarrow D^{+} D^{-}}  =  \frac{1}{2 M}\cdot
2 \cdot \frac{1}{16 \pi } \frac{2 |{\bm q}_1|}{E_{cm}}     \cdot
\left|
i \mathcal{M}_{\Phi\rightarrow D^{+} D^{-}}(P,q)
\right|^2  ,
\ea


\section{Numerical Results}

\subsection{Dependence of binding energy $E_b$ on the energy scales $\Lambda$ and $\Delta$ defined in the F.F.}

Before the numerical computations, we have simplified the tensor indices by
using the package of ``{\rm FeynCalc 6.0.0}¡±\cite{FeynCalc} on the platform of ``{\rm Wolfram Mathematica}¡±.

By taking $\Delta$ at some fixed values, i.e., $\Delta = m_\pi,2m_\pi,3m_\pi,4m_\pi,5m_\pi,4\pi f_\pi, m_1 +m_2, +\infty$, we have taken a search for the qualified BSWF solutions
in the range of $0.8\,GeV \leq \Lambda  \leq 4.0\,GeV $ and $-150\,MeV  \leq  E_b < 0$.
The so-called qualified BSWF solutions $\Phi(|{\bm p}|)$ are those without zero values at $|{\bm p}|<+\infty$ since the Schrodinger wavefunctions $\Phi(|{\bm p}|)$ in (\ref{SchWF-VS-BSWF}) would have no zeros for a regular Hamiltonian $H=H(|{\bm p}|)$ according to the SchE of form $H \Phi=E\Phi $ in the momentum space.
It is shown that,
for the bound states of both $D^{\ast}\bar{D}^{\ast}$ and $B^{\ast}\bar{B}^{\ast}$ with spin quantum numbers of $J=0$:  \\
\noindent (1) in the $\Delta = \{ m_\pi,2m_\pi,3m_\pi,4m_\pi,5m_\pi \}$ cases:
in the $I=0$ case there can exist qualified BSWF solutions,
while in the $I=1$ case there cannot exist qualified BSWF solutions; \\
\noindent (2) in the $\Delta = \{ 4\pi f_\pi, m_1 +m_2, +\infty \}$ cases,
neither in the $I=0$ case nor in the $I=1$ case
there can exist qualified BSWF solutions.
This results in the $\Delta = +\infty$ case are different with results in the SchE formalism\cite{SchE-scheme}\cite{single-term-bad-Lagrangian}, that may be caused by the approximation taken for the polarization vectors, $\xi(p) \rightarrow 1$, mentioned in the paragraph above Eq. (\ref{define-NewExp-factor-2}).

For the $J=0, I=1$ case, our results are opposite to the ones in Ref.\cite{BSE-DstarDstar-KeHongWei}. That might be caused by the different definitions of the BS wavefunctions. In our definition (\ref{BSWF-J0-A}), there are three independent wavefunctions while there is only one in Ref.\cite{BSE-DstarDstar-KeHongWei}, so there would be more constraint on the
qualification of the wavefunctions, that means, the qualified BSE solutions in Ref.\cite{BSE-DstarDstar-KeHongWei} might be unqualified in our BSE (\ref{BSE-pq-J0I0-d4q}).

\begin{figure}[!htbp]
\centering
\includegraphics[scale=0.8]{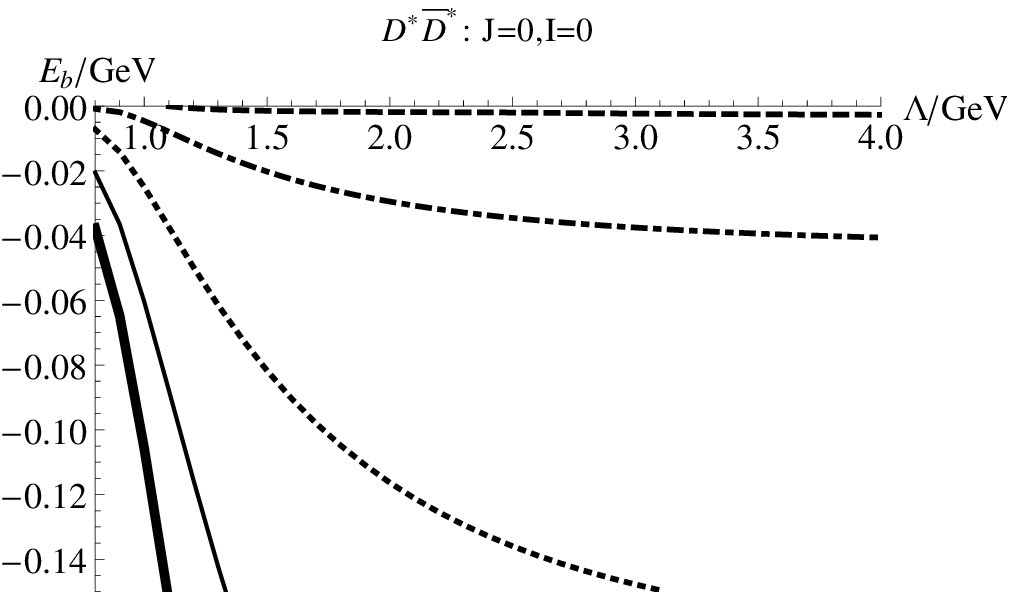}\\
\centering (a)\\
\par
\centering
\includegraphics[scale=0.99]{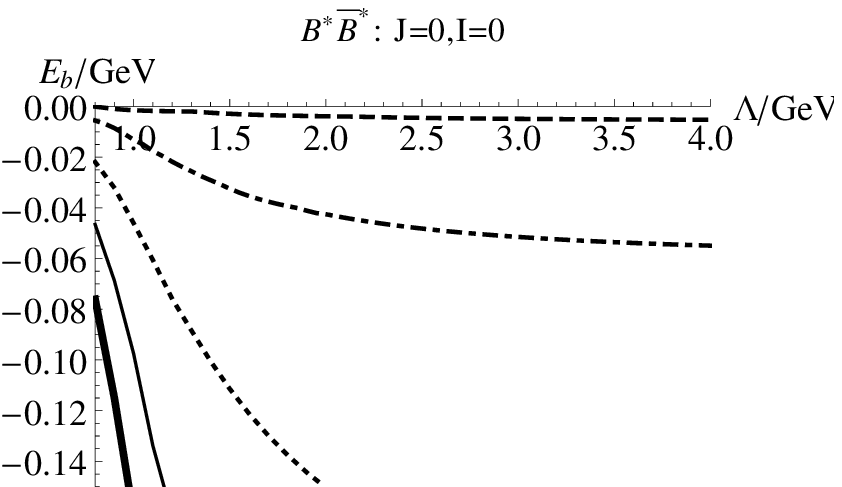}\\
\centering (b)\\
\caption{
Dependence of binding energy $E_b$ on the energy scales $\Delta$ and $\Lambda$ defined in (\ref{FF-dot-Exp})
for the bound states of (a) $D^{\ast}\bar{D}^{\ast}$ and (b) $B^{\ast}\bar{B}^{\ast}$  with $J=0$ and $I=0$, where from top to bottom,
the dashed, dotdashed, dotted, thin solid, thick solid lines are for the $\Delta= m_\pi,2m_\pi,3m_\pi,4m_\pi,5m_\pi$ cases, respectively.
There doesn't exist qualified BSWF solutions for the bound states of $D^{\ast}\bar{D}^{\ast}$ and $B^{\ast}\bar{B}^{\ast}$ with $J=0$ and $I=1$.
}
\label{J0I0-rank64x3-Eb+BSWF}
\end{figure}


\subsection{Normalized BS wavefunctions}

\begin{figure}[!htbp]
\noindent
\includegraphics[scale=0.9]{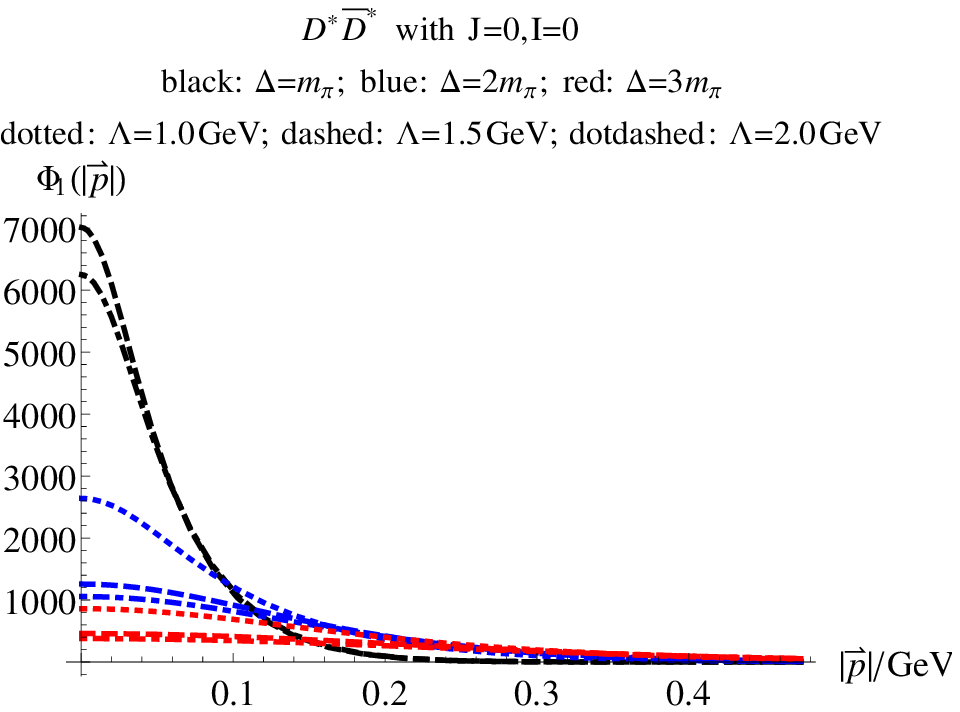}\\
\noindent
\includegraphics[scale=0.9]{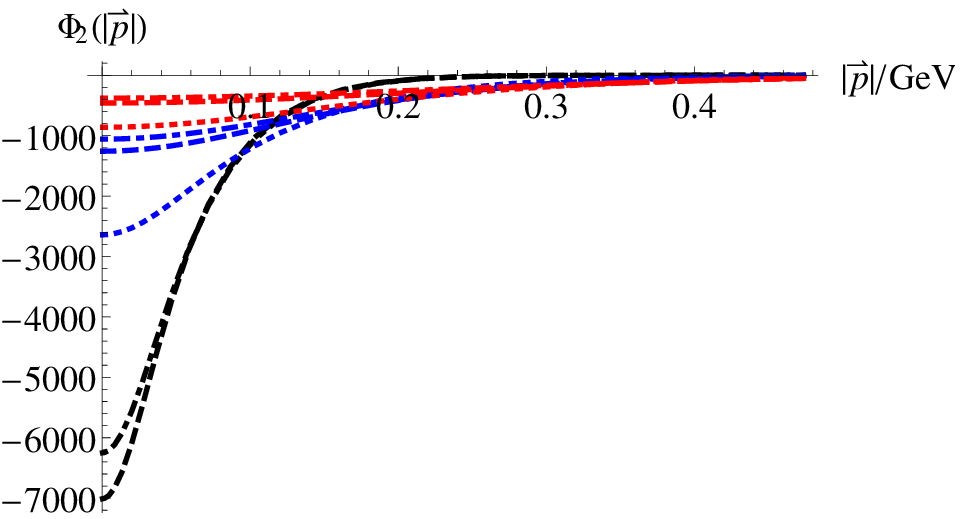}\\
\noindent
\includegraphics[scale=0.9]{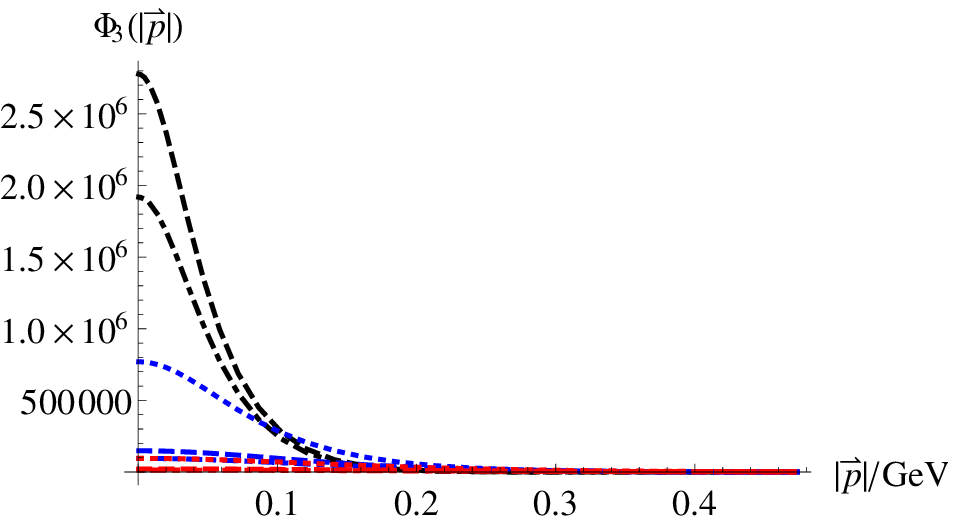}
\caption{
The BSWFs  $\Phi_{1,2,3}(|{\bm p}|)$ of $D^{\ast}\bar{D}^{\ast}$ bound states with $J=0$ and $I=0$, where
the two black, three blue, three red lines are for $\Delta=m_\pi, 2 m_\pi, 3 m_\pi$, respectively. In each case,
the dotted, dashed and dot-dashed lines are for $\Lambda=1.0,1.5,2.0 \,GeV$, respectively.
}
\label{J0I0-DstarDstar-plot-Phi123}
\end{figure}

\begin{figure}[!htbp]
\noindent
\includegraphics[scale=0.9]{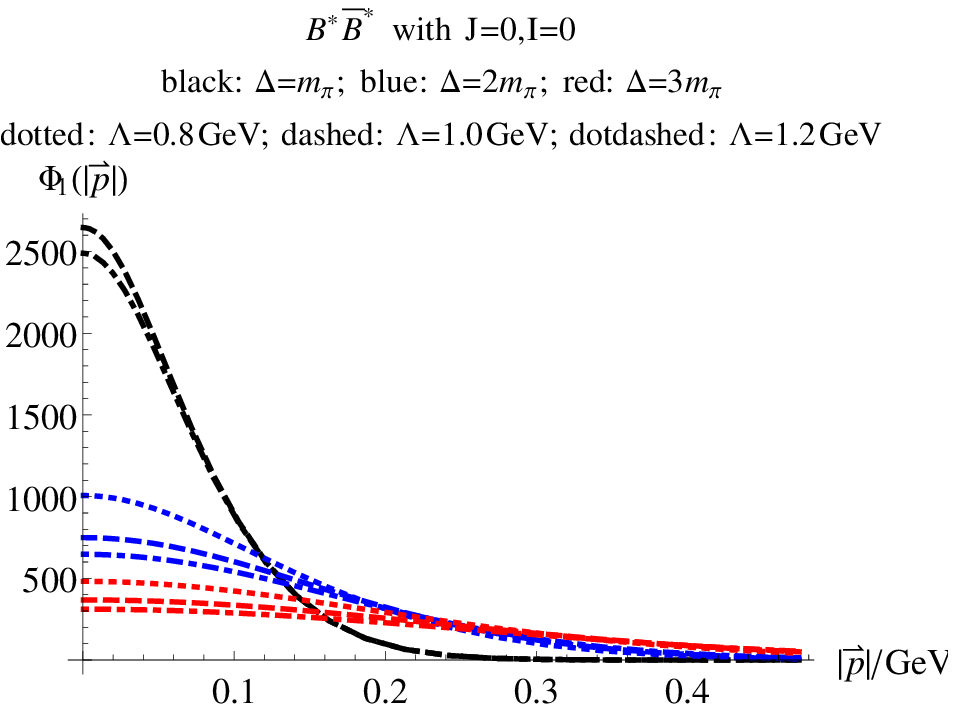}\\
\includegraphics[scale=0.9]{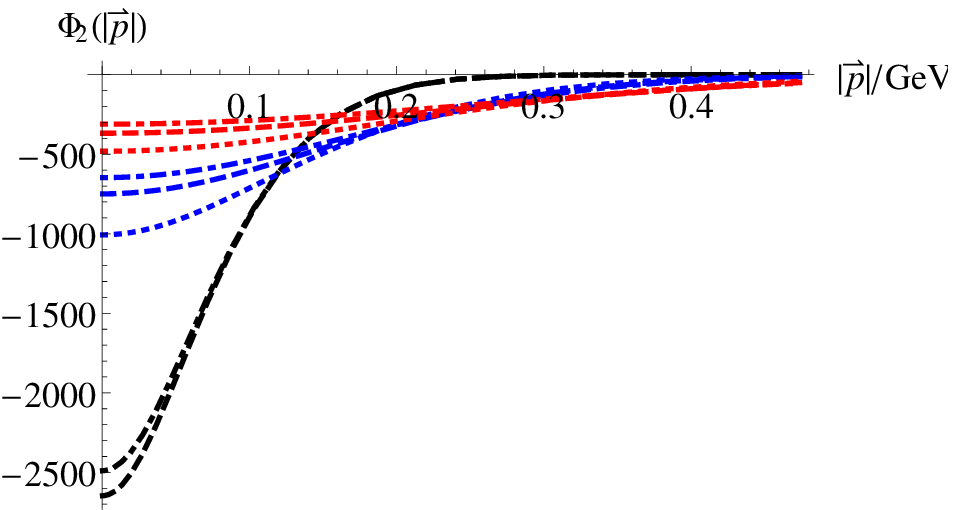}\\
\includegraphics[scale=0.9]{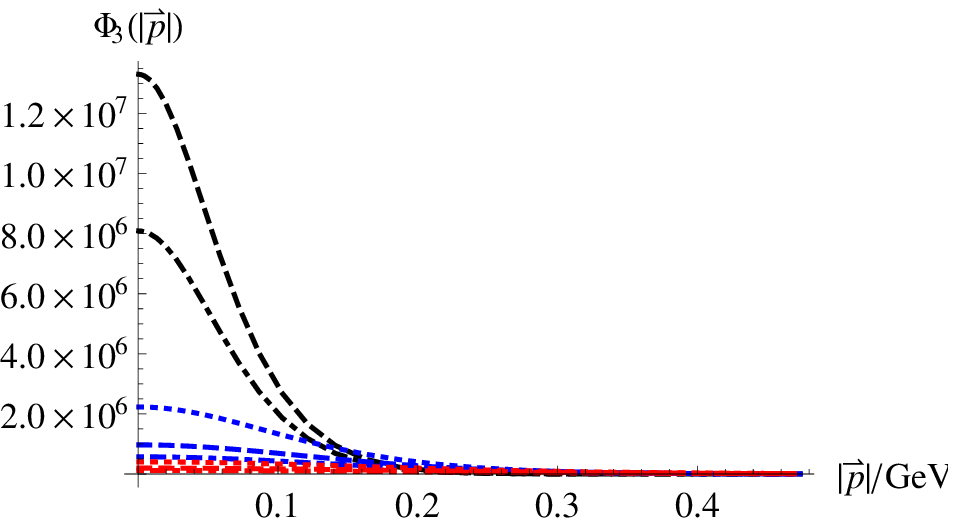}
\caption{
The BSWFs  $\Phi_{1,2,3}(|{\bm p}|)$ of $B^{\ast}\bar{B}^{\ast}$ bound states with $J=0$ and $I=0$, where
the two black, three blue, three red lines are for $\Delta=m_\pi, 2 m_\pi, 3 m_\pi$, respectively. In each case,
the dotted, dashed and dot-dashed lines are for $\Lambda=0.8,1.0,1.2 \,GeV$, respectively.
}
\label{J0I0-BstarBstar-plot-Phi123}
\end{figure}

The normalized BS wavefunctions are necessary for calculating the decay widths of the bound states, and they were plotted in Fig. \ref{J0I0-DstarDstar-plot-Phi123} for $D^{\ast}\bar{D}^{\ast}$ systems and Fig. \ref{J0I0-BstarBstar-plot-Phi123} for $B^{\ast}\bar{B}^{\ast}$ systems.

Indeed, we can get some useful results even without a normalization of the BSWF, such as: \\
(1) we can check the reasonability or the self-consistency of the $\Delta$ values we have used in solving the BSE, according to the line shape of the BS wavefunctions in Fig.~\ref{J0I0-DstarDstar-plot-Phi123} and Fig.~\ref{J0I0-BstarBstar-plot-Phi123}£»
and we can know $\Delta$ is really of the order of $m_{\pi}$ and is obviously smaller than $\Lambda_\chi = 4 \pi f_{\pi}$, which is consistent with our results above, i.e., there doesn't exist qualified BSWF solutions in the $\Delta = 4\pi f_\pi, m_1 +m_2, +\infty$ cases;\\
(2) the results of $\Phi_1 \simeq -\Phi_2$ (but $\Phi_1 \neq -\Phi_2$) are good enough to support the reasonability of a requirements in Ref. \cite{Sanchis-Alepuz-glueball}, i.e., only space-like components of $\chi_{P}^{\mu\nu}$ surviving, or, the reasonability of a choice of a Coulomb gauge in Ref. \cite{Cui-glueball}; and this is straightforward by reminding that
we will have $P_\mu \chi_{P}^{\mu\nu}(p_t^2) = P_\nu \chi_{P}^{\mu\nu}(p_t^2) \simeq 0$ with $P \cdot p_t =0$ and $\Phi_1 \simeq -\Phi_2$ in the center-of-mass frame. \\

\subsection{Decay width}

Due to the mass difference $2m_{D^{\ast+}}-2m_{D^{+}}
= 0.28122 \,GeV > 0$,
the decay process $D^{\ast}\bar{D}^{\ast} \rightarrow D^{+}D^{-} $ is allowed by the kinetic phase space.
For $X(3940)$, due to the mass difference
$2m_{D^{\ast+}} -m_{X(3940)}
= 0.08052 \,GeV > 0$,
although its $J^{PC}=?^{??}$ has not been determined, its mass ($\simeq 3942 \,MeV$) and total decay width ($< 52 \,MeV$) has been determined;
it is allowed to be a candidate of the $0^{++}$ molecular states of $D^{\ast}\bar{D}^{\ast}$.\\

Due to the mass difference $2m_{B^{\ast+}}-2m_{B^{+}}
= 0.0908 \,GeV >0$,
the decay process $B^{\ast}\bar{B}^{\ast} \rightarrow  B^{-} \bar{B}^{0}$ is also allowed by the kinetic phase space. Besides,
if the bounding energy $|E_b|>0.0908 \,GeV=90.8 \,MeV$, then the decay width of $B^{\ast}\bar{B}^{\ast}$ might be narrow since the decay $B^{\ast}\bar{B}^{\ast}\rightarrow B \bar{B} $ is forbidden by the kinetic phase space.


\begin{table}[!htbp]
\caption{\label{table-decay-DstarDstar-J0I0-Nmpion}
Decay widths of the $D^{\ast} \bar{D}^{\ast} $ bound states for different $\Delta$  and $\Lambda$  values.}
\begin{tabular}{c|cccc}
\hline
\hline
$\Delta = m_{\pi}$ & & & & \\
\hline
 $\Lambda(GeV)$  & $1.5$  &  $2.0$  &  $2.5$   & $3.5$ \\
\hline
 $E_b(MeV)$   & $ - 1.5 $ &  $- 1.9$   & $ - 2.0 $    &  $ - 2.6 $ \\
\hline
 $\Gamma(\Phi\rightarrow D^+ D^-) (MeV)$    & $2.69 \times 10^{-24}$ &  $  2.37 \times 10^{-26} $   &  $2.11 \times 10^{-24} $ &   $ 9.45 \times 10^{-24} $   \\
\hline
\end{tabular}
\begin{tabular}{c|ccccc}
\hline
$\Delta = 2 m_{\pi}$ &  &  &  &  &   \\
\hline
 $\Lambda(GeV)$    &   $1.0$   &   $1.5$    &   $2.0$  &   $2.5$    & $3.5$  \\
\hline
 $E_b(MeV)$    &   $- 4.6$    &  $- 20.2 $    & $- 29.5$   &  $- 34.6$     &  $- 39.4$   \\
\hline
 $\Gamma(\Phi\rightarrow D^+ D^-) (MeV)$   & $9.13 \times 10^{-5}$    & $6.73 \times 10^{-5}$   &   $2.87\times 10^{-6} $   & $1.84 \times 10^{-5}$      & $ 1.46 \times 10^{-4 }$  \\
\hline
\end{tabular}
\begin{tabular}{c|ccc}
\hline
$\Delta = 3 m_{\pi}$ &  &  &  \\
\hline
 $\Lambda(GeV)$  & $1.0$    &  $1.5$   & $2.0$ \\
\hline
 $E_b(MeV)$   & $ - 25.0 $    & $ - 81.7 $    &  $ - 116.3 $ \\
\hline
 $\Gamma(\Phi\rightarrow D^+ D^-) (MeV)$    &  $  0.836 $ & $  1.328  $  & $ 1.228 $   \\
\hline
\hline
\end{tabular}
\end{table}


\begin{table}[!htbp]
\caption{\label{table-decay-BstarBstar-J0I0-Nmpion}
Decay widths of the $B^{\ast} \bar{B}^{\ast} $ bound states for different $\Delta$  and $\Lambda$ values.}
\begin{tabular}{c|cccccc}
$\Delta = m_{\pi}$  &   &   &   &   &   &   \\
\hline
 $\Lambda(GeV)$    &   $0.8$   &   $1.0$   &   $1.2$   &   $1.5$    &   $2.5$    & $3.5$  \\
\hline
 $E_b(MeV)$     & $ - 0.07 $ & $- 1.5$ & $- 1.9 $  &  $- 2.9$     &  $- 4.5$     & $ - 5.1$   \\
\hline
 $\Gamma(\Phi\rightarrow B^{-} \bar{B}^{0}) (MeV)$   & $1 \times 10^{-20}$ & $ 6\times 10^{-20} $ & $6 \times 10^{-20}$  &  $5 \times 10^{-20}$     &  $ 2 \times 10^{-22} $     & $3 \times 10^{-20}$  \\
\end{tabular}
\begin{tabular}{c|ccccccc}
$\Delta = 2 m_{\pi}$  &   &   &   &   &   &   &   \\
\hline
 $\Lambda(GeV)$   &  $0.8$   &  $1.0$  &  $1.2$ &  $1.5$ &  $2.0$   &   $2.5$    & $3.5$  \\
\hline
 $E_b(MeV)$    &  $- 5.6$ & $- 13.1$  &  $ - 21.8$ & $- 32.5 $ & $ - 42.5$ &  $- 48.2$     &   $- 53.5$  \\
\hline
 $\Gamma(\Phi\rightarrow  B^{-} \bar{B}^{0}) (MeV)$   &  $0.003   $ & $ 0.013 $   &  $  0.037 $  & $0.113$  &   $ 0.266$  & $ 0.434$     &   $ 0.707 $  \\
\end{tabular}
\begin{tabular}{c|ccc}
$\Delta = 3 m_{\pi}$  &  &  &  \\
\hline
 $\Lambda(GeV)$    & $0.8$    &  $1.0$   & $1.2$ \\
\hline
 $E_b(MeV)$   &  $- 21.9  $     &  $ - 46.0$     & $ - 75.8$   \\
\hline
 $\Gamma(\Phi\rightarrow B^{-} \bar{B}^{0}) (MeV)$    &  $  16.538  $     &  $  97.726  $     & $  361.864 $  \\
\end{tabular}
\end{table}

By comparing the results in the cases of $\Delta = m_{\pi}$,
$\Delta = 2 m_{\pi}$
and $\Delta = 3 m_{\pi}$,
(see Table \ref{table-decay-DstarDstar-J0I0-Nmpion} and Table \ref{table-decay-BstarBstar-J0I0-Nmpion})
we can see that the values of the banding energy and the widths of $D^{\ast}\bar{D}^{\ast}$ or $B^{\ast}\bar{B}^{\ast}$ systems are very sensitive to the value of $\Delta$, and that might be resulted by the complicate expressions of the total interaction kernel ${\ovl K}^{total}$ in (\ref{kernel-J0I0}), see  Appedndix \ref{appendix-kernel-in-BSE}.


\section{Summary}

We study the scalar bound states of $D^{\ast}\bar{D}^{\ast}$ and $B^{\ast}\bar{B}^{\ast}$ in the
Bethe-Salpeter (BS) formalism,
with the effective interaction kernel
extracted
from the chiral perturbative theory (ChPT) and the heavy quark effective theory (HQET)
in the ladder approximation and the covariant instantaneous approximation.
The results show that, in the scalar case ($J=0$), there can only exist $I = 0$ bound states for parameters in proper range,
while there cannot exist the $I = 1$ bound states in the whole reasonable parameter range.
It is partly because of that, there are more constraints on the BS wavefunction if there is more than one independent function in the Lorentz structure of the BS wavefunction.

\section{Acknowledgements}

The author is very grateful to
Prof. Xin-Heng GUO at Beijing Normal University,
Dr. Jia-Jun WU at University of Chinese Academy of Sciences
and
Dr. Xing-Hua WU at Yulin Normal University,
for the important and essential guidance.

\newpage

\appendix

\section{Effective Kernel in the BSE}\label{appendix-kernel-in-BSE}

With
keeping in mind that
the factor
$$\mathcal{F}(k,\Lambda,m_\phi;p,q,\Delta)
\equiv [F(k,\Lambda,m_\phi)]^2 \cdot [\mathbb{F}(|{\bm p}|,|{\bm q}|,\Delta)]^2 $$
defined in (\ref{FF-dot-Exp})
and the instantaneous approximation
should be complemented
in the following
calculation,
here we list $\overline{K}^{(\phi) 11,11} _{\sigma\gamma\nu\beta} (P,p,q)$ and $\overline{K}^{(\phi) 11,22} _{\sigma\gamma\nu\beta} (P,p,q)$ as below (with $\phi = \pi,\eta,\rho,\omega,\sigma$ denoting the exchanged light-flavor mesons): \\

(1) $\pi$-exchange\\
\noindent
\ba
&& \overline{K}^{(\pi) 11,11} _{\sigma\gamma\nu\beta} (P,p,q) \nonumber\\
 &=&  (2\pi)^4 \delta^{(4)}(p_1+p_2 -q_1-q_2)  \nonumber\\
 &&\cdot
  i( i\frac{g_{\pi}}{f_{\pi}})\epsilon^{\alpha_1  \mu_1 \nu_1 \lambda_1} (2 p_1  - k)_{\alpha_1} g_{\mu_1 \sigma}g_{\lambda_1 \nu}(- \frac{1}{\sqrt{2}} )  (ik_{\nu_1})
\nonumber\\
 &&\cdot   \frac{i}{ k^2 - m_{\pi}^2}  \nonumber\\
 &&\cdot
   i(i\frac{g_{\pi}}{f_{\pi}})\epsilon^{\alpha_2
     \mu_2 \nu_2 \lambda_2}  (2 p_2 + k)_{\alpha_2} g_{\mu_2 \beta}g_{\lambda_2 \gamma}(- \frac{1}{\sqrt{2}} ) (-i      k_{\nu_2})
,\label{Kpion1111}
\ea

\ba
&& \overline{K}^{(\pi) 11,22} _{\sigma\gamma\nu\beta} (P,p,q) \nonumber\\
 &=&(2\pi)^4 \delta^{(4)}(p_1+p_2 -q_1-q_2)  \nonumber\\
 &&\cdot
  i(i\frac{g_{\pi}}{f_{\pi}})\epsilon^{\alpha_1
     \mu_1 \nu_1 \lambda_1} (2 p_1  - k)_{\alpha_1}g_{\mu_1 \sigma}g_{\lambda_1 \nu} \cdot  1  \cdot  (ik_{\nu_1})
\nonumber\\
 &&\cdot  \frac{i}{   k^2 - m_{\pi}^2} \nonumber\\
 &&\cdot
   i(i\frac{g_{\pi}}{f_{\pi}})\epsilon^{\alpha_2\mu_2 \nu_2 \lambda_2} (2 p_2 + k)_{\alpha_2}g_{\mu_2 \beta}g_{\lambda_2 \gamma}  \cdot 1  \cdot
(-ik_{\nu_2})
,
\ea
or
\ba
  {\ovl K}^{(\pi)11,22}_{\sigma\gamma\nu\beta} = 2 {\ovl K}^{(\pi)11,11}_{\sigma\gamma\nu\beta},\,\label{Kpi1122VSKpi1111}
\ea

(2) $\eta$-exchange\\
\ba
&&  \overline{K}^{(\eta) 11,11} _{\sigma\gamma\nu\beta}  (P,p,q) \nonumber\\
 &=&(2\pi)^4 \delta^{(4)}(p_1+p_2 -q_1-q_2)  \nonumber\\
 &&\cdot
  i(i\frac{g_{\pi}}{f_{\pi}})\epsilon^{\alpha_1
     \mu_1 \nu_1 \lambda_1}  (2 p_1  - k)_{\alpha_1} g_{\mu_1 \sigma}g_{\lambda_1 \nu}(\frac{1}{\sqrt{6}}) (ik_{\nu_1})
\nonumber\\
 &&\cdot
 \frac{i}{ k^2 - m_{\eta}^2} \nonumber\\
 &&\cdot
   i(i\frac{g_{\pi}}{f_{\pi}})\epsilon^{\alpha_2
     \mu_2 \nu_2 \lambda_2}  (2 p_2  + k)_{\alpha_2} g_{\mu_2 \beta}g_{\lambda_2 \gamma}(\frac{1}{\sqrt{ 6}}) (-ik_{\nu_2})
,
\ea

\ba
\overline{K}^{(\eta) 11,22}_{\sigma\gamma\nu\beta} = 0  ,\label{Keta1122VSKeta1111}
\ea

(3) $\rho$-exchange\\
\ba
&&\overline{K}^{(\rho) 11,11}_{\sigma\gamma\nu\beta} (P,p,q) \nonumber\\
 &=&(2\pi)^4 \delta^{(4)}(p_1+p_2 -q_1-q_2)  \nonumber\\
 &&\cdot
  i\left[ \frac{\sqrt{2}}{2} g_{\beta}g_{V}  \cdot
      g_{\mu_1 \sigma}g^{\mu_1}_{\nu}  (2 p_1  - k)_{\alpha_1} (-\frac{1}{\sqrt{2}}) g^{\alpha_1 \chi}    \right.
\nonumber\\
 && \left.    - i 2 \sqrt{2} \overline{M}_{12} g_{\lambda} g_{V}\cdot
      g_{\mu_1 \sigma}g_{\nu_1 \nu}(- \frac{1}{\sqrt{2}} ) (ik^{\mu_1}g^{\nu_1 \chi} -    ik^{\nu_1}g^{\mu_1 \chi})
    \right]
\nonumber\\
 &&\cdot
    \frac{ -i(g_{\chi \phi} -  k_{\chi}k_{\phi} /m_{\rho}^2) }{   k^2 - m_{\rho}^2 }  \nonumber\\
 &&\cdot
   i \left[(-\frac{\sqrt{2}}{2}  g_{\beta} g_{V})\cdot
      g_{\mu_2 \gamma}g^{\mu_2}_{ \beta}  (2 p_2  + k)_{\alpha_2}(-\frac{1}{\sqrt{2}}) g^{\alpha_2 \phi}        \right.
\nonumber\\
 &&  \left.   + i \cdot 2 \sqrt{2}  \overline{M}_{12} g_{\lambda}g_{V} \cdot
      g_{\mu_2 \gamma}g_{\nu_2 \beta}(-\frac{1}{\sqrt{2}}) (-ik^{\mu_2}g^{\nu_2 \phi} +      ik^{\nu_2}g^{\mu_2 \phi} )
    \right],
\ea

\ba
&& \overline{K}^{(\rho) 11,22}_{\sigma\gamma\nu\beta} (P,p,q) \nonumber\\
 &=&(2\pi)^4 \delta^{(4)}(p_1+p_2 -q_1-q_2)  \nonumber\\
 &&\cdot
  i \left[ \frac{\sqrt{2}}{2} g_\beta g_{V} \cdot
     g_{\mu_1 \sigma}g^{\mu_1}_{ \nu}  (2 p_1  - k)_{\alpha_1} \cdot 1 \cdot     g^{\alpha_1 \chi}  \right.
\nonumber\\
 &&\left.
     -i 2 \sqrt{2} \overline{M}_{12} g_{\lambda} g_{V} \cdot
     g_{\mu_1 \sigma}g_{\nu_1 \nu} \cdot 1 \cdot (ik^{\mu_1}g^{\nu_1 \chi} -        ik^{\nu_1}g^{\mu_1 \chi})     \right]
  \nonumber\\
 &&\cdot
    \frac{ -i(g_{\chi \phi} -  k_{\chi}k_{\phi} /m_{\rho}^2) }{   k^2 - m_{\rho}^2 }  \nonumber\\
 &&\cdot
   i \left[(-\frac{\sqrt{2}}{2} g_\beta g_V) \cdot
     g_{\mu_2 \gamma}g^{\mu_2}_{ \beta}  (2 p_2  + k)_{\alpha_2}  \cdot  1  \cdot      g^{\alpha_2 \phi}   \right.
\nonumber\\
 && \left.
     +i 2 \sqrt{2}  \overline{M}_{12}  g_{\lambda} g_{V} \cdot      g_{\mu_2 \gamma}g_{\nu_2 \beta}  \cdot  1  \cdot  (-i k^{\mu_2}g^{\nu_2 \phi} +   ik^{\nu_2}g^{\mu_2 \phi} )
    \right],
\ea
or
\ba
 {\ovl K}^{(\rho)11,22}_{\sigma\gamma\nu\beta} = 2 {\ovl K}^{(\rho)11,11}_{\sigma\gamma\nu\beta},\,\label{Krho1122VSKrho1111}
\ea

(4) $\omega$-exchange\\
\ba
&&\overline{K}^{(\omega) 11,11}_{\sigma\gamma\nu\beta} (P,p,q) \nonumber\\
 &=&(2\pi)^4 \delta^{(4)}(p_1+p_2 -q_1-q_2)  \nonumber\\
 &&\cdot
  i \left[ \frac{\sqrt{2}}{2}g_{\beta}g_{V} \cdot
      g_{\mu_1 \sigma}g^{\mu_1}_{ \nu}  (2 p_1  - k)_{\alpha_1}\frac{1}{\sqrt{2}}    g^{\alpha_1 \chi}   \right.
\nonumber\\
 &&\left.
     - i2 \sqrt{2} \overline{M}_{12} g_{\lambda}g_{V} \cdot
      g_{\mu_1 \sigma}g_{\nu_1 \nu}\frac{1}{\sqrt{2}} (ik^{\mu_1}g^{\nu_1 \chi} -   ik^{\nu_1}g^{\mu_1 \chi})      \right]
\nonumber\\
 &&\cdot
  \frac{-i(g_{\chi \phi} - k_{\chi}k_{\phi}/m_{\omega}^2)}{   k^2 - m_{\omega}^2 }  \nonumber\\
 &&\cdot
   i \left[(- \frac{\sqrt{2}}{2} g_{\beta}g_{V})\cdot
      g_{\mu_2 \gamma}g_{\mu_2 \beta}  (2 p_2  + k)_{\alpha_2}\frac{1}{\sqrt{2}} g^{\alpha_2 \phi}   \right.
\nonumber\\
 &&\left.
     + i 2 \sqrt{2}  \overline{M}_{12} g_{\lambda}g_{V} \cdot
      g_{\mu_2 \gamma}g_{\nu_2 \beta}\frac{1}{\sqrt{2}} (-ik^{\mu_2}g^{\nu_2 \phi} +    ik^{\nu_2}g^{\mu_2 \phi} )      \right],
\ea

\ba
\overline{K}^{(\omega) 11,22}_{\sigma\gamma\nu\beta}= 0  ,\label{Komega1122VSKomega1111}
\ea

(5) $\sigma$-exchange\\
\ba
 \overline{K}^{(\sigma) 11,11} _{\sigma\gamma\nu\beta} (P,p,q)
&=&(2\pi)^4 \delta^{(4)}(p_1+p_2 -q_1-q_2)  \nonumber\\
 &&\cdot
  i\cdot 2  \overline{M}_{12} g_s  \cdot  g_{\mu_1 \sigma}g^{\mu_1}_{\nu} \cdot \frac{i}{ k^2 - m_{\sigma}^2} \cdot  i \cdot  2  \overline{M}_{12} g_s  \cdot
   g_{\mu_2 \beta}g^{\mu_2}_{\gamma}   ,
\ea

\ba
\overline{K}^{(\sigma) 11,22}_{\sigma\gamma\nu\beta} = 0,\label{Ksigma1122VSKsigma1111}
\ea

\noindent so that,  the total kernel will be:

\ba
{\ovl K}^{total}_{\sigma\gamma\nu\beta}
&=&\left[{\ovl K}^{11,11}+{\ovl K}^{11,22}\right]_{\sigma\gamma\nu\beta} \quad   \mbox{( for $I=0$ case)}\nonumber\\
&=&[({\ovl K}^{(\pi)11,11}+{\ovl K}^{(\pi)11,22})
+({\ovl K}^{(\eta)11,11}+{\ovl K}^{(\eta)11,22})\nonumber\\
&&+({\ovl K}^{(\rho)11,11}+{\ovl K}^{(\rho)11,22})
+({\ovl K}^{(\omega)11,11}+{\ovl K}^{(\omega)11,22})\nonumber\\
&&+({\ovl K}^{(\sigma)11,11}+{\ovl K}^{(\sigma)11,22})]_{\sigma\gamma\nu\beta} \nonumber\\
&=&[3 {\ovl K}^{(\pi)11,11}
+ {\ovl K}^{(\eta)11,11}
+ 3 {\ovl K}^{(\rho)11,11}
+ {\ovl K}^{(\omega)11,11}
+ {\ovl K}^{(\sigma)11,11} ]_{\sigma\gamma\nu\beta}  ;\nonumber\\
\ea
or
\ba
{\ovl K}^{total}_{\sigma\gamma\nu\beta}
&=&\left[{\ovl K}^{11,11}-{\ovl K}^{11,22}\right]_{\sigma\gamma\nu\beta}  \quad  \mbox{( for $I=1$ case)}\nonumber\\
&=&[({\ovl K}^{(\pi)11,11}-{\ovl K}^{(\pi)11,22})
+({\ovl K}^{(\eta)11,11}-{\ovl K}^{(\eta)11,22})\nonumber\\
&&+({\ovl K}^{(\rho)11,11}-{\ovl K}^{(\rho)11,22})
+({\ovl K}^{(\omega)11,11}-{\ovl K}^{(\omega)11,22})\nonumber\\
&&+({\ovl K}^{(\sigma)11,11}-{\ovl K}^{(\sigma)11,22})]_{\sigma\gamma\nu\beta} \nonumber\\
&=&[ -{\ovl K}^{(\pi)11,11}
+ {\ovl K}^{(\eta)11,11}
-{\ovl K}^{(\rho)11,11}
+ {\ovl K}^{(\omega)11,11}
+ {\ovl K}^{(\sigma)11,11} ]_{\sigma\gamma\nu\beta} .\nonumber\\
\ea


\section{Effective Kernel in the Decay Process}\label{appendix-kernel-in-decay}

With
keeping in mind that
the factor
$$\mathcal{F}(k,\Lambda,m_\phi;p,q,\Delta)
\equiv [F(k,\Lambda,m_\phi)]^2 \cdot [\mathbb{F}(|{\bm p}|,|{\bm q}|,\Delta)]^2$$
 defined in (\ref{FF-dot-Exp})
but without the instantaneous approximation
should be complemented
in the following
calculation,
here we list $K_{\alpha\beta}^{(\pi,\eta,\rho,\omega) ij,i'j'}(P,p,q)$ as below: \\

(1) $\pi$-exchange\\

\ba
&& K^{(\pi) 11,11}  _{\alpha\beta} (P,p,q) \nonumber\\
 &=&
  i(-\frac{2g_\pi}{f_\pi}\bar{M})  (-\frac{1}{\sqrt{2}}) (  i\cdot 2k^{\alpha})
\non\\
&&\cdot   \frac{i}{ k^2 - m_{\pi}^2} \non\\
&&\cdot
i( \frac{2g_\pi}{f_\pi}\bar{M})    (-\frac{1}{\sqrt{2}}) (- i\cdot 2k_{\beta}) ,
\ea

\ba
&& K^{(\pi) 11,22}  _{\alpha\beta} (P,p,q) \nonumber\\
 &=&
  i(-\frac{2g_\pi}{f_\pi}\bar{M})  \cdot 1 \cdot (  i\cdot 2k^{\alpha})
\non\\
&&\cdot   \frac{i}{ k^2 - m_{\pi}^2} \non\\
&&\cdot
i( \frac{2g_\pi}{f_\pi}\bar{M})     \cdot 1 \cdot (- i\cdot 2k_{\beta}) ,
\ea
or
\ba
K^{(\pi) 11,22}  =  2   K^{(\pi) 11,11} ;
\ea

\ba
&& K^{(\pi) 22,11} _{\alpha\beta} (P,p,q) \nonumber\\
 &=&
  i(-\frac{2g_\pi}{f_\pi}\bar{M})  \cdot 1\cdot (i\cdot 2k_{\alpha})
\non\\
&&
\cdot   \frac{i}{ k^2 - m_{\pi}^2} \non\\
&&\cdot
i( \frac{2g_\pi}{f_\pi}\bar{M})   \cdot 1\cdot  (-i\cdot 2k_{\beta}) ,
\ea
or
\ba
 K^{(\pi)22,11} = 2 K^{(\pi)11,11} ,
\ea

\ba
&& K^{(\pi) 22,22} _{\alpha\beta} (P,p,q) \nonumber\\
 &=&
  i(-\frac{2g_\pi}{f_\pi}\bar{M})  \cdot (\frac{1}{\sqrt{2}})\cdot (i\cdot 2k_{\alpha})
\non\\
&&
\cdot   \frac{i}{ k^2 - m_{\pi}^2} \non\\
&&\cdot
i( \frac{2g_\pi}{f_\pi}\bar{M})   \cdot (\frac{1}{\sqrt{2}}) \cdot  (-i\cdot 2k_{\beta}) ,
\ea
or
\ba
 K^{(\pi) 22,22}  =  \frac{1}{2}  K^{(\pi) 22,11}   =  \frac{1}{2} \cdot (2 K^{(\pi)11,11})
=    K^{(\pi)11,11} ;
\ea
As said above, due to the isospin  conservation, we only need consider the $I'=I,\,I'_3=I_3$ case.
For convenience, when we choose the $I'_3=I_3=0$ channel, we can mark $C_{(I',I'_3)}^{i'j'}$ and $C_{(I,I_3)}^{ij}$ to $C_{(I)}^{i'j'}$ and $C_{(I)}^{ij}$ by dropping the index $I'_3$ and $I_3$, and we can have
\ba
{\mathcal K}^{(\pi)}_{+- \rightarrow +-}
&=&  C_{I}^{+-}  \cdot K^{(\pi) 11,11}  \cdot C_{I}^{+-},\\
{\mathcal K}^{(\pi)}_{+- \rightarrow 00}
&=& C_{I}^{+-}  \cdot K^{(\pi) 11,22}  \cdot C_{I}^{00}
=C_{I}^{+-}  \cdot 2 K^{(\pi) 11,11}  \cdot  C_{I}^{00} ,\\
{\mathcal K}^{(\pi)}_{00 \rightarrow +-}
&=& C_{I}^{00}  \cdot K^{(\pi) 22,11}  \cdot C_{I}^{+-}
=C_{I}^{00}  \cdot 2 K^{(\pi)11,11}   \cdot   C_{I}^{+-},\\
{\mathcal K}^{(\pi)}_{00 \rightarrow 00}
&=& C_{I}^{00}  \cdot K^{(\pi) 22,22}  \cdot C_{I}^{00}
=C_{I}^{00}  \cdot  K^{(\pi)11,11} \cdot C_{I}^{00},
\ea
and consequently
\ba
&&{\mathcal K}^{(\pi)}_{+- \rightarrow +-}
+{\mathcal K}^{(\pi)}_{+- \rightarrow 00}
+{\mathcal K}^{(\pi)}_{00 \rightarrow +-}
+{\mathcal K}^{(\pi)}_{00 \rightarrow 00} \non\\
&=&
K^{(\pi) 11,11}  \cdot
\left(
C_{I}^{+-}  \cdot C_{I}^{+-}
+ 2  C_{I}^{+-}    \cdot  C_{I}^{00}
+ 2  C_{I}^{00}   \cdot C_{I}^{+-}
+C_{I}^{00}   \cdot C_{I}^{00}
\right),
\ea

(2) $\eta$-exchange\\

\ba
&& K^{(\eta) 11,11} _{\alpha\beta}  (P,p,q) \nonumber\\
 &=&
  i(-\frac{2g_\pi}{f_\pi}\bar{M})   (\frac{1}{\sqrt{6}}) (i\cdot 2k_{\alpha})
\non\\
&&
\cdot   \frac{i}{ k^2 - m_{\eta}^2} \non\\
&&\cdot
i( \frac{2g_\pi}{f_\pi}\bar{M})    (\frac{1}{\sqrt{6}}) (-i\cdot 2k_{\beta}) ,
\ea

\ba
K^{(\eta) 11,22} _{\alpha\beta}  (P,p,q) =0,
\ea

\ba
K^{(\eta) 22,11} _{\alpha\beta}  (P,p,q)= 0,
\ea

\ba
&& K^{(\eta) 22,22} _{\alpha\beta}  (P,p,q) \nonumber\\
 &=&
  i(-\frac{2g_\pi}{f_\pi}\bar{M})   (\frac{1}{\sqrt{6}}) (i\cdot 2k_{\alpha})
\non\\
&&
\cdot   \frac{i}{ k^2 - m_{\eta}^2} \non\\
&&\cdot
i( \frac{2g_\pi}{f_\pi}\bar{M})    (\frac{1}{\sqrt{6}}) (-i\cdot 2k_{\beta}) ,
\ea
or
\ba
K^{(\eta) 22,22} =  K^{(\eta) 11,11} ;
\ea

\ba
{\mathcal K}^{(\eta)}_{+- \rightarrow +-} &=& C_{I}^{+-}  \cdot K^{(\eta) 11,11}  \cdot C_{I}^{+-},\\
{\mathcal K}^{(\eta)}_{+- \rightarrow 00} &=& C_{I}^{+-}  \cdot K^{(\eta) 11,22}  \cdot C_{I}^{00} =0 ,\\
{\mathcal K}^{(\eta)}_{00 \rightarrow +-} &=& C_{I}^{00}  \cdot K^{(\eta) 22,11}  \cdot C_{I}^{+-}=0,\\
{\mathcal K}^{(\eta)}_{00 \rightarrow 00} &=& C_{I}^{00}  \cdot K^{(\eta) 22,22}  \cdot C_{I}^{00}=C_{I}^{00}  \cdot  K^{(\eta)11,11} \cdot C_{I}^{00},
\ea
and consequently
\ba
&&{\mathcal K}^{(\eta)}_{+- \rightarrow +-}
+{\mathcal K}^{(\eta)}_{+- \rightarrow 00}
+{\mathcal K}^{(\eta)}_{00 \rightarrow +-}
+{\mathcal K}^{(\eta)}_{00 \rightarrow 00} \non\\
&=&
 K^{(\eta) 11,11}  \cdot
\left(
 C_{I}^{+-}  \cdot C_{I}^{+-}
+C_{I}^{00}   \cdot C_{I}^{00}
\right);
\ea

(3) $\rho$-exchange\\
\ba
&&K^{(\rho) 11,11}_{\alpha\beta} (P,p,q) \nonumber\\
&=&
i (i\sqrt{2}\lambda g_v\epsilon_{\lambda_1\alpha_1\beta_1\mu_1} ) g^{\mu_1\alpha} (i \cdot 2 (-q_1-p_1)^{\lambda_1})
(-\frac{1}{\sqrt{2}})  (i k^{\alpha_1}g^{\beta_1\nu})
\nonumber\\
&&\cdot
\frac{ -i(g_{\nu \beta} -  k_{\nu}k_{\beta} /m_{\rho}^2) }{   k^2 - m_{\rho}^2 }  \nonumber\\
&&\cdot
i (i\sqrt{2}\lambda g_v\epsilon_{\lambda_2\alpha_2\beta_2\mu_2} ) g^{\mu_2\beta} (i \cdot 2 (-q_2-p_2)^{\lambda_2})
(-\frac{1}{\sqrt{2}}) (-i k^{\alpha_2} g^{\beta_2\beta})  ,
\ea

\ba
&&K^{(\rho) 11,22}_{\alpha\beta} (P,p,q) \nonumber\\
&=&
i (i\sqrt{2}\lambda g_v\epsilon_{\lambda_1\alpha_1\beta_1\mu_1} ) g^{\mu_1\alpha} (i \cdot 2 (-q_1-p_1)^{\lambda_1})
\cdot 1 \cdot  (i k^{\alpha_1}g^{\beta_1\nu})
\nonumber\\
&&\cdot
\frac{ -i(g_{\nu \beta} -  k_{\nu}k_{\beta} /m_{\rho}^2) }{   k^2 - m_{\rho}^2 }  \nonumber\\
&&\cdot
i (i\sqrt{2}\lambda g_v\epsilon_{\lambda_2\alpha_2\beta_2\mu_2} ) g^{\mu_2\beta} (i \cdot 2 (-q_2-p_2)^{\lambda_2})
\cdot 1 \cdot  (-i k^{\alpha_2} g^{\beta_2\beta})  ,
\ea
or
\ba
K^{(\rho) 11,22}  =  2   K^{(\rho) 11,11};
\ea

\ba
&&K^{(\rho) 22,11}_{\alpha\beta} (P,p,q) \nonumber\\
&=&
i (i\sqrt{2}\lambda g_v\epsilon_{\lambda_1\alpha_1\beta_1\mu_1} ) g^{\mu_1\alpha} (i \cdot 2 (-q_1-p_1)^{\lambda_1})
\cdot 1 \cdot   (i k^{\alpha_1}g^{\beta_1\nu})
\nonumber\\
&&\cdot
\frac{ -i(g_{\nu \beta} -  k_{\nu}k_{\beta} /m_{\rho}^2) }{   k^2 - m_{\rho}^2 }  \nonumber\\
&&\cdot
i (i\sqrt{2}\lambda g_v\epsilon_{\lambda_2\alpha_2\beta_2\mu_2} ) g^{\mu_2\beta} (i \cdot 2 (-q_2-p_2)^{\lambda_2})
\cdot 1 \cdot  (-i k^{\alpha_2} g^{\beta_2\beta})  ,
\ea
or
\ba
 K^{(\rho)22,11}  = 2 K^{(\rho)11,11} ,\,
\ea

\ba
&&K^{(\rho) 22,22}_{\alpha\beta} (P,p,q) \nonumber\\
&=&
i (i\sqrt{2}\lambda g_v\epsilon_{\lambda_1\alpha_1\beta_1\mu_1} ) g^{\mu_1\alpha} (i \cdot 2 (-q_1-p_1)^{\lambda_1})
\cdot \frac{1}{\sqrt{2}} \cdot   (i k^{\alpha_1}g^{\beta_1\nu})
\nonumber\\
&&\cdot
\frac{ -i(g_{\nu \beta} -  k_{\nu}k_{\beta} /m_{\rho}^2) }{   k^2 - m_{\rho}^2 }  \nonumber\\
&&\cdot
i (i\sqrt{2}\lambda g_v\epsilon_{\lambda_2\alpha_2\beta_2\mu_2} ) g^{\mu_2\beta} (i \cdot 2 (-q_2-p_2)^{\lambda_2})
\cdot  \frac{1}{\sqrt{2}} \cdot  (-i k^{\alpha_2} g^{\beta_2\beta})  ,
\ea
or
\ba
K^{(\rho) 22,22}  =  \frac{1}{2}    K^{(\rho) 22,11}    =  \frac{1}{2}  \cdot  2 K^{(\rho)11,11}  =    K^{(\rho)11,11} £»
\ea

\ba
{\mathcal K}^{(\rho)}_{+- \rightarrow +-} &=&C_{I}^{+-}  \cdot K^{(\rho) 11,11}  \cdot C_{I}^{+-},\\
{\mathcal K}^{(\rho)}_{+- \rightarrow 00} &=&C_{I}^{+-}  \cdot K^{(\rho) 11,22}  \cdot C_{I}^{00}=C_{I}^{+-}  \cdot  2   K^{(\rho) 11,11}\cdot   C_{I}^{00}  ,\\
{\mathcal K}^{(\rho)}_{00 \rightarrow +-} &=&C_{I}^{00}  \cdot K^{(\rho) 22,11}  \cdot C_{I}^{+-}= C_{I}^{00}  \cdot 2   K^{(\rho)11,11} \cdot  C_{I}^{+-} ,\\
{\mathcal K}^{(\rho)}_{00 \rightarrow 00} &=&C_{I}^{00}  \cdot K^{(\rho) 22,22}  \cdot C_{I}^{00}= C_{I}^{00}  \cdot     K^{(\rho)11,11} \cdot C_{I}^{00} ,
\ea
and consequently
\ba
&&{\mathcal K}^{(\rho)}_{+- \rightarrow +-}
+{\mathcal K}^{(\rho)}_{+- \rightarrow 00}
+{\mathcal K}^{(\rho)}_{00 \rightarrow +-}
+{\mathcal K}^{(\rho)}_{00 \rightarrow 00} \non\\
&=&
 K^{(\rho) 11,11}  \cdot
\left(
C_{I}^{+-}   \cdot C_{I}^{+-}
+2   C_{I}^{+-}  \cdot    C_{I}^{00}
+2   C_{I}^{00}   \cdot  C_{I}^{+-}
+ C_{I}^{00}   \cdot C_{I}^{00}
\right);
\ea

(4) $\omega$-exchange\\

\ba
&&K^{(\omega) 11,11}_{\alpha\beta} (P,p,q) \nonumber\\
&=&
i (i\sqrt{2}\lambda g_v\epsilon_{\lambda_1\alpha_1\beta_1\mu_1} ) g^{\mu_1\alpha} (i \cdot 2 (-q_1-p_1)^{\lambda_1})
 (\frac{1}{\sqrt{2}})  (i k^{\alpha_1}   g^{\beta_1\nu})
\nonumber\\
&&\cdot
\frac{ -i(g_{\nu \beta} -  k_{\nu}k_{\beta} /m_{\omega}^2) }{   k^2 - m_{\omega}^2 }  \nonumber\\
&&\cdot
i (i\sqrt{2}\lambda g_v\epsilon_{\lambda_2\alpha_2\beta_2\mu_2} ) g^{\mu_2\beta} (i \cdot 2 (-q_2-p_2)^{\lambda_2})
 (\frac{1}{\sqrt{2}}) (-i k^{\alpha_2} g^{\beta_2\beta})  ,
\ea

\ba
K^{(\omega) 11,22}_{\alpha\beta} (P,p,q)  =0,
\ea

\ba
K^{(\omega) 22,11}_{\alpha\beta} (P,p,q)
=0,
\ea

\ba
&&K^{(\omega) 22,22}_{\alpha\beta} (P,p,q) \nonumber\\
&=&
i (i\sqrt{2}\lambda g_v\epsilon_{\lambda_1\alpha_1\beta_1\mu_1} ) g^{\mu_1\alpha} (i \cdot 2 (-q_1-p_1)^{\lambda_1})
 (\frac{1}{\sqrt{2}})  (i k^{\alpha_1}   g^{\beta_1\nu})
\nonumber\\
&&\cdot
\frac{ -i(g_{\nu \beta} -  k_{\nu}k_{\beta} /m_{\omega}^2) }{   k^2 - m_{\omega}^2 }  \nonumber\\
&&\cdot
i (i\sqrt{2}\lambda g_v\epsilon_{\lambda_2\alpha_2\beta_2\mu_2} ) g^{\mu_2\beta} (i \cdot 2 (-q_2-p_2)^{\lambda_2})
 (\frac{1}{\sqrt{2}}) (-i k^{\alpha_2} g^{\beta_2\beta})  ,
\ea
or
\ba
K^{(\omega) 22,22}  =  K^{(\omega) 11,11};
\ea

\ba
{\mathcal K}^{(\omega)}_{+- \rightarrow +-} &=&C_{I}^{+-}  \cdot K^{(\omega) 11,11}  \cdot C_{I}^{+-},\\
{\mathcal K}^{(\omega)}_{+- \rightarrow 00} &=&C_{I}^{+-}  \cdot K^{(\omega) 11,22}  \cdot C_{I}^{00}= 0 ,\\
{\mathcal K}^{(\omega)}_{00 \rightarrow +-} &=&C_{I}^{00}  \cdot K^{(\omega) 22,11}  \cdot C_{I}^{+-}= 0  ,\\
{\mathcal K}^{(\omega)}_{00 \rightarrow 00} &=&C_{I}^{00}  \cdot K^{(\omega) 22,22}  \cdot C_{I}^{00}= C_{I}^{00}  \cdot   K^{(\omega) 11,11} \cdot C_{I}^{00}  ,
\ea
and consequently
\ba
&&{\mathcal K}^{(\omega)}_{+- \rightarrow +-}
+{\mathcal K}^{(\omega)}_{+- \rightarrow 00}
+{\mathcal K}^{(\omega)}_{00 \rightarrow +-}
+{\mathcal K}^{(\omega)}_{00 \rightarrow 00} \non\\
&=&
 K^{(\omega) 11,11}  \cdot
\left(
C_{I}^{+-}     \cdot C_{I}^{+-}
+ C_{I}^{00}    \cdot C_{I}^{00}
\right);
\ea


\newpage


\end{document}